\documentclass[aps,prl,10pt,twocolumn,groupedaddress,longbibliography]{revtex4-1}
\usepackage{amsmath}
\usepackage{graphicx}
\usepackage{color}

\begin{document}

\title{Electrical generation of surface plasmon polaritons in plasmonic heterostructures}

\author{Maxim Trushin}
\email{mxt@nus.edu.sg}
\affiliation{Department of Materials Science and Engineering and Institute for Functional Intelligent Materials, National University of Singapore, Singapore}

\date{\today}

\begin{abstract}
Surface plasmon polaritons (SPPs) can be understood as two-dimensional light confined to a conductor-dielectric interface via plasmonic excitations.
While low-energy SPPs behave similarly to photons, higher-frequency SPPs resemble surface plasmons.
Electrically generating mid-range SPPs is particularly challenging because it requires compensating for momentum mismatch,
a process conventionally achieved through inelastic electron transport in nanostructures.
Here, we theoretically demonstrate that electrical SPP generation is possible
by directly coupling electron-hole dipoles to the quantized SPP field across an insulating spacer without accompanying electron transport.
This approach can be realized in plasmonic van der Waals heterostructures composed of strongly-biased monolayer graphene as the emitter, few-layer hexagonal boron nitride as the spacer, and silver (or gold) as the plasmonic material.
In this configuration, graphene's remarkable ability to support a strongly non-equilibrium steady-state electron-hole population results in non-thermal, bias-tunable SPP emission that is uniform along the hBN/Ag interface, achieving a power conversion efficiency of up to 1\% and a Purcell factor of up to 100. 
These findings pave the way for integrating photonic and electronic functionalities within a single two-dimensional heterostructure.
\end{abstract}

\maketitle

{\it Introduction. ---}
Light-matter interactions are central to optoelectronic devices, enabling energy conversion between photons and electrons \cite{rosencher2002optoelectronics}.
While confining electrons to two-dimensional (2D) planes at semiconductor interfaces (such as GaAs/AlGaAs) has led to devices with exceptionally high carrier mobility, achieving similar confinement for photons remains challenging. Surface plasmon polaritons (SPPs), quasiparticles formed by photons coupled to surface plasmons
at a metal-dielectric interface \cite{zayats2005nano,maier2007plasmonics}, offer a promising solution \cite{ozbay2006-photonics-electronics}.
Conventional SPP generation methods require sub-wavelength 
nanostructuring \cite{PRL1968grating,ebbesen1998extraordinary,kravets2018plasmonic,echtermeyer2016surface,rawashdeh2023high,PRL2024,PRL2025}, nonlinear wave mixing \cite{palomba2008nonlinear,constant2016all}, or complex electrical approaches involving either keV electron beams
\cite{PRB1975keVbeam,Electron-beam-review2019,generation2023keVbeam,generation2024keVbeam} or nanoantenna fabrication \cite{PRL2016antenna,zhang2019antenna,PRL2020antenna,zhang2021electrical,pommier2023nanoscale,wang2023engineering} to facilitate inelastic electron tunneling \cite{lambe1976light,PRL1996localexcitation,bharadwaj2011electrical,cui2020electrically,fung2020too,du2017highly,zhu2020hot}. However, these techniques typically produce SPPs only in highly localized regions, such as beneath scanning tunneling microscope tips \cite{wang2011excitation,fei-rodin2012gate,zhang2013edge,ni2018fundamental}, as predicted in earlier theoretical work \cite{PRL-exp-th-1991,johansson1990theory,uehara1992theory}. 
Moreover, strong absorption at the interface
\cite{bell1973surface,alexander1974dispersion} demands either aggressive miniaturization of photonic components  or active compensation for propagation losses \cite{SSPemissionPRL2008,SSPemissionNanoLett2008,SPPamplification2010,Review2012}.
Recent experiments have used van der Waals heterostructures \cite{tunneling-th2017plasmon,wang2021optical,wang2023exciton,wang2024upconversion} for SPP generation via inelastic tunneling, while new theoretical proposals suggest coupling surface plasmons with spin waves in magnetic structures to achieve low-frequency excitations \cite{costa2023strongly,yuan2024breaking}. In this Letter, we present an alternative concept for electrical generation of SPPs.

The concept builds on van der Waals heterostructures \cite{geim2013van,novoselov2016review},
which provide a unique platform for creating {\em two} interacting interfaces separated by an atomically thin,
electrically insulating spacer. In this design, one interface confines 2D electron-hole dipoles that emit SPPs,
while the other serves as the SPP-supporting interface. Graphene is the optimal choice for the emitter material,
as it can sustain a high bias that generates a strongly non-equilibrium electron-hole population \cite{above2000K-2017hBN}.
The plasmonic interface requires a low-loss noble metal film \cite{mironov2024graphene,yakubovsky2023optical}
(such as Au or Ag) and a dielectric layer that also functions as the insulating spacer \cite{illarionov2020insulators}. Few-layer hexagonal boron nitride (hBN) is ideal for this purpose, provided it is carefully engineered: it must be thin enough to enable strong coupling between the SPP field and electron-hole dipoles for efficient SPP generation, yet thick enough to suppress electron tunneling from graphene to the underlying metal. Additionally, hBN encapsulation enhances charge carrier mobility in graphene, which is crucial for maintaining a strongly non-equilibrium electron-hole population under high bias \cite{drift2017hBN}.

The proposed setup differs fundamentally from previous plasmonic van der Waals heterostructures \cite{wang2021optical,wang2023exciton,wang2024upconversion}, 
where SPP generation relies on inelastic electron tunneling. In our design, neither electrons nor holes are expected to tunnel through the hBN barrier.
Instead, electron-hole dipoles couple to SPPs via the SPP field itself, which pierces graphene across the hBN spacer.
Because the SPP field penetration length greatly exceeds the spacer thickness, the resulting coupling between electron-hole dipoles and SPPs is anticipated to be much stronger than what can be achieved through quantum mechanical tunneling. A steady-state electron-hole population in biased graphene ensures continuous SPP emission.
Furthermore, current fabrication techniques enable the production of high-quality graphene/hBN heterostructures, allowing for uniform SPP generation in central regions far from the edges.

\begin{figure*}
 \includegraphics[width=\textwidth]{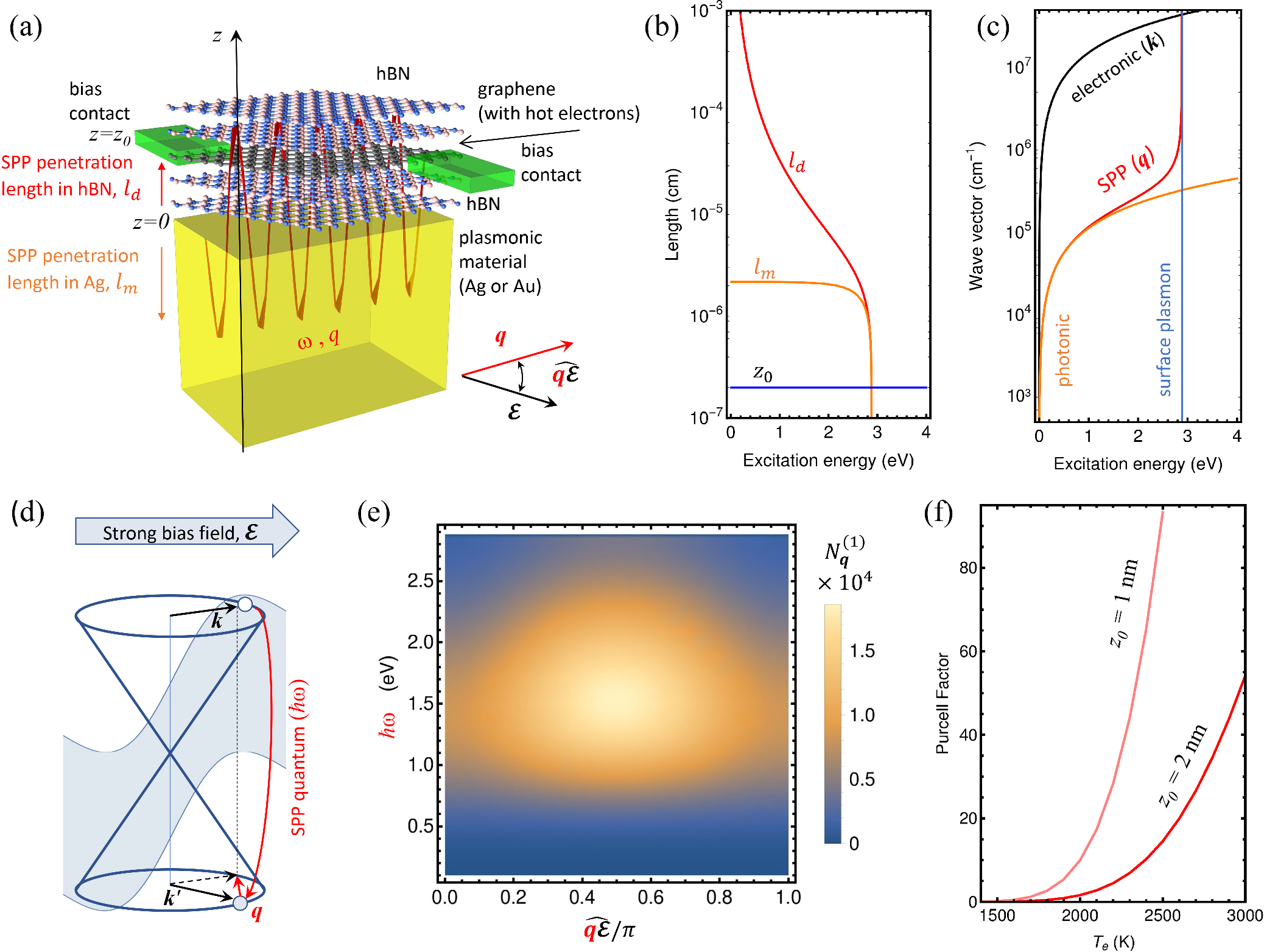}
 \caption{SPP generation by biased electrons in an hBN/graphene/hBN/Ag heterostructure.
 (a) Schematic of the proposed device, where highly mobile graphene electrons, driven out of equilibrium by a strong bias field, generate SPPs at the Ag/hBN interface. 
 (b) SPP field penetration lengths in the dielectric ($l_d$) and metal ($l_m$),
 compared to the separation $z_0$ between the Ag/hBN interface and the graphene layer.
 (c) Dispersion relations for graphene electrons, photons in hBN, SPPs, and surface plasmons,
 illustrating a momentum mismatch in the near-infrared excitation range. 
 (d) SPP emission via electron-hole recombination in strongly biased graphene, where the bias induces carrier population inversion in both energy and momentum, as indicated by the shaded electron bands.
 Due to the inequality $q\ll |\mathbf{k}'-\mathbf{k}|$, SPPs are emitted nearly normal to the electron motion under bias.
 (e) Angle-resolved non-equilibrium SPP distribution function at $T_e\sim 2800$K and nearly equal drift and band velocities, 
 showing a preferred propagation direction perpendicular to the bias field, with a peak in the near-infrared region ($\sim$1.5 eV). 
 (f) The Purcell factor, defined as the ratio of SPP to photon emission rates, demonstrates enhancement of 2D light emission at $T_e>1500$K
 and $z_0<2$nm.
 \label{fig1}
 }
\end{figure*}

{\it Model. ---}
Figure \ref{fig1}(a) illustrates the setup, where SPP propagates along the interface between hBN and silver
characterized respectively by the dielectric permittivity $\epsilon_d=4.97$ \cite{hBN2018dielectric} and metallic permittivity function
$\epsilon_m = \epsilon_\infty - {\omega_p^2}/{\omega^2}$,
where $\hbar\omega$ is the excitation energy, $\hbar \omega_p=9.1$ eV is the plasmon energy quanta \cite{silver2008}, $\hbar$ is the Planck constant,
and  $\epsilon_\infty=5$ is the high-frequency permittivity limit \cite{yang2015optical}.
A more accurate representation involving the $\omega$-dependence and anisotropy of $\epsilon_d$ 
is important only at low frequencies \cite{PRL2015tomadin} and for bulk emission \cite{fang2024hyperbolic}, 
respectively, while the focus of this work is on near-infrared emission in a 2D limit.
The SPP field exponentially vanishes in silver and hBN at the scale of the penetration lengths $l_m$ and $l_d$ shown in Fig. \ref{fig1}(b).
Silver can be substituted by gold \cite{olmon2012optical} or another low-loss conductor \cite{ziyatkhan2024quest} without qualitative changes incurred.
The 2D electrons in graphene and SPPs are separated by the distance $z_0$ (the thickness of a few hBN layers) much smaller than $l_d$
allowing the SPP field to couple with electrons. 
Figure  \ref{fig1}(c) shows the dispersion relations for electrons in graphene
$E_{\pm k}=\pm \hbar v k$ ($v=10^8$ cm/s, $k$ is the electron wave vector), and for propagating SPPs 
with the SPP wave vector $q$ given by
\begin{equation}
 q = \frac{\omega}{c} \sqrt{\frac{\epsilon_d \left(\frac{\omega_p^2}{\omega^2}-\epsilon_\infty\right)}
 {\frac{\omega_p^2}{\omega^2}-\epsilon_\infty - \epsilon_d}},
 \label{SPPk}
\end{equation}
where $c$ is the speed of light.
The SPP energy quanta is limited from above by the surface plasmon frequency $\omega_{sp}=\omega_p/\sqrt{\epsilon_\infty+\epsilon_d}$
(about $2.9$ eV for Ag/hBN interface), and it follows photonic dispersion $q\sim \sqrt{\epsilon_d}\omega/c$ in low-frequency limit.

SPP excitation requires both excess energy and momentum, which can be supplied by electrons in strongly biased graphene.
Graphene's exceptional ability to sustain high bias fields leads to electron temperatures exceeding 2000 K in high-quality hBN-encapsulated samples \cite{above2000K-2017hBN},
with drift velocities approaching the band velocity in undoped graphene \cite{drift2011theory} on hBN \cite{drift2017hBN}
and even SiO$_2$ \cite{drift2010SiO2}.
At such high temperatures, electron-hole pairs form in the conduction and valence bands and are accelerated by the bias field, creating a strongly non-equilibrium distribution, as shown in Fig. \ref{fig1}(d). The characteristic electron and hole momenta correspond to the extrema of their respective distribution functions in the momentum space.

Recombination occurs via indirect interband transitions, emitting SPP quanta into the Ag/hBN interface. Since the near-infrared SPP momentum is much smaller than the electron and hole momenta, the emitted SPP wave vector is nearly perpendicular to both. Because electron and hole wave vectors align with the electric field, the angle-resolved non-equilibrium SPP distribution in Fig. \ref{fig1}(e) exhibits a clear maximum when the SPP wave vector is perpendicular to the field. SPP emission is non-thermal, with its peak position and intensity governed by the bias field through electron temperature and drift velocity.
The emission enhancement, as compared to the photon emission into vacuum, is characterized by the Purcell factor shown in Fig. \ref{fig1}(f).
The results indicate that the proposed setup is viable for practical applications \cite{qian2021spontaneous} provided  $T_e\gtrsim 2000$K and $z_0\lesssim 2$nm.

\begin{figure}
 \includegraphics[width=\columnwidth]{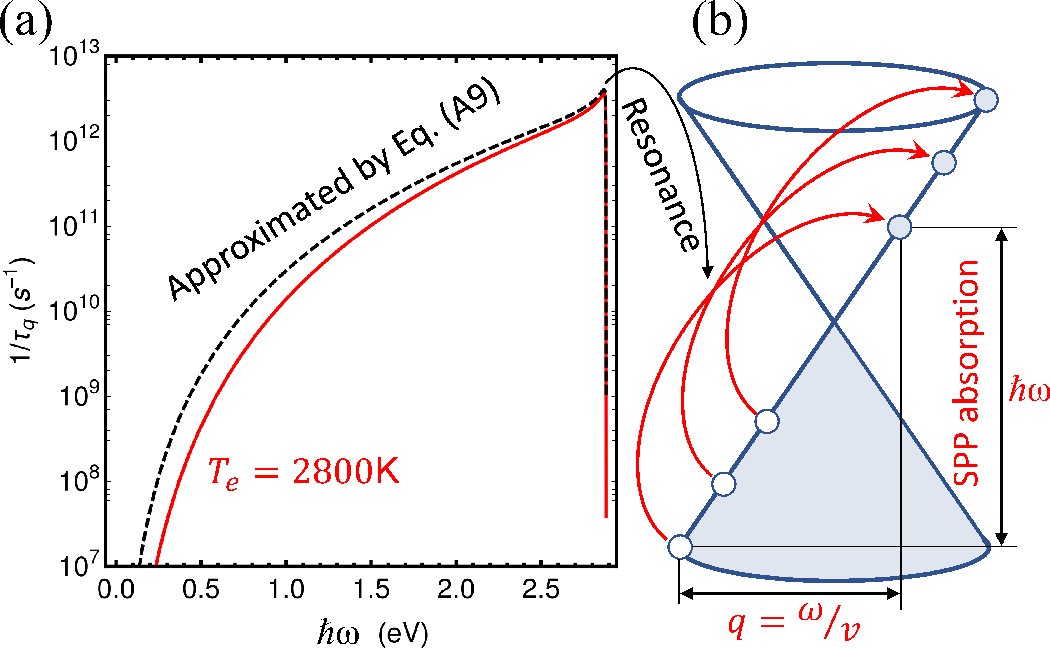}
 \caption{Extrinsic SPP attenuation due to electrons in graphene.
 (a) SPP decay rate in graphene due to SPP absorption by hot electrons at $T_e=2800$K.
 The dashed curve corresponds to the approximate model with the SPP decay rate given by Eq. (\ref{tauq}).
 (b) Extrinsic SPP decay rate rapidly approaches the intrinsic (interfacial) values of the order of $10^{13}$ s$^{-1}$ at the resonance 
  $vq=\omega$ due to a nesting effect in the SPP absorption.
 \label{fig2}
 }
\end{figure}

\begin{figure}
 \includegraphics[width=0.93\columnwidth]{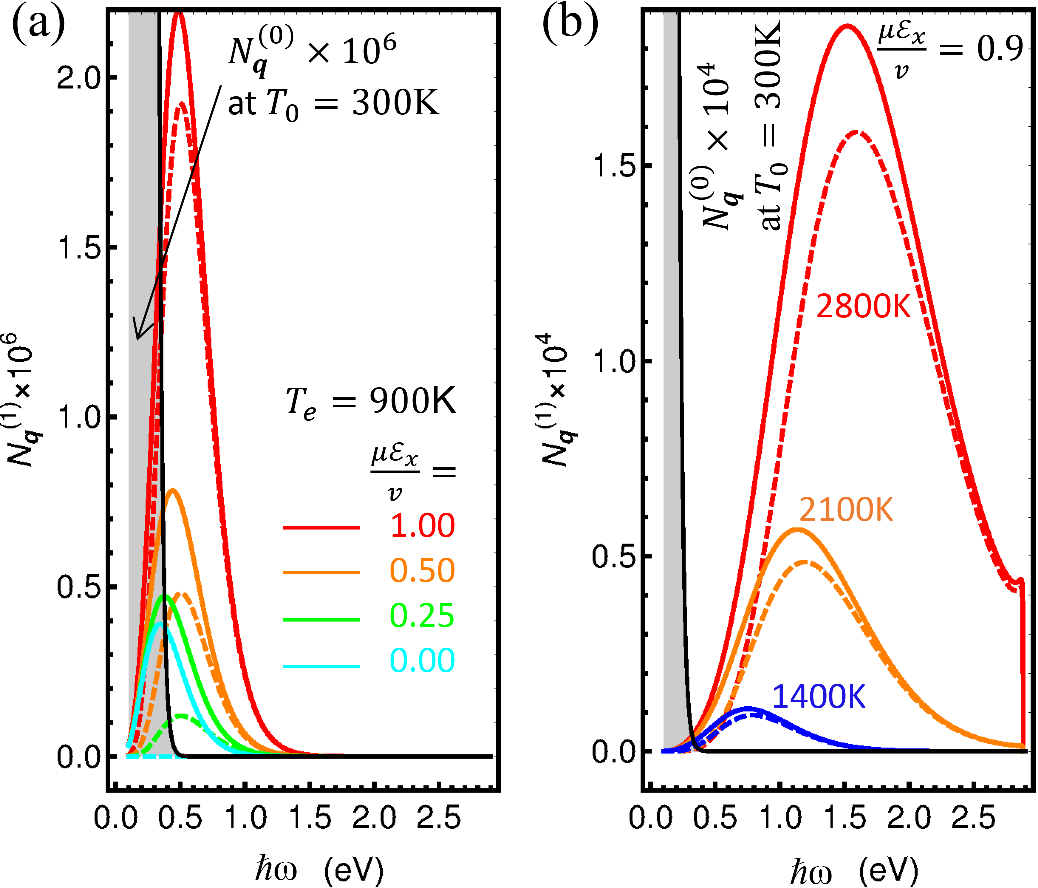}
 \caption{Non-equilibrium SPP distribution function compared to the thermal SPP bath distribution (gray shaded area). Solid curves represent the full solution, while dashed curves correspond to the approximate model,
 where the SPP emission is due to non-thermalized electrons only and the generation rate is given by Eq. (\ref{Gq}). (a) The peak of the propagating SPP distribution shifts away from the thermally excited background as the electron drift velocity approaches the graphene band velocity under increasing bias. (b) At strong bias fields, the non-equilibrium SPP distribution becomes more pronounced with rising electron temperature relative to the SPP bath.
 The maximum emission occurs at the energy estimated by Eq. (\ref{spp-max}). Note the different scale of the ordinate axis in panels (a,b).
 \label{fig3}
 }
\end{figure}

{\it Results. ---}
SPPs are evolving in quantum mechanical emission \cite{fakonas2014two,tame2013quantum,ballester2010quantum}
described by the Hamiltonian \cite{archambault2010quantum,bernardi2015theory} explicitly given in the End Matter.
The Hamiltonian describes electrons and SPPs by means of the respective occupations, $n_{\pm \mathbf{k}}$ and $N_\mathbf{q}$,
similar to electron-phonon scattering \cite{phillips2012advanced}. The SPP emission/absorption transitions can be written as
\begin{eqnarray}
\nonumber && |n_{-(\mathbf{k}-\mathbf{q})},n_{+\mathbf{k}};N_\mathbf{q}\rangle \xrightarrow{emi}
|n_{-(\mathbf{k}-\mathbf{q})}+1,n_{+\mathbf{k}}-1;N_\mathbf{q}+1 \rangle, \\
\nonumber && 
|n_{+(\mathbf{k}+\mathbf{q})},n_{-\mathbf{k}};N_\mathbf{q}\rangle \xrightarrow{abs}
|n_{+(\mathbf{k}+\mathbf{q})}+1,n_{-\mathbf{k}}-1;N_\mathbf{q}-1 \rangle.\\
\label{transit}
\end{eqnarray}
The electron occupation is given by the sum $n_{\pm\mathbf{k}}=f_{\pm\mathbf{k}}^{(0)}+f_{\pm\mathbf{k}}^{(1)}$
where $f_{\pm\mathbf{k}}^{(0)}$ is the Fermi-Dirac distribution at the temperature $T_e$,
and $f_{\pm\mathbf{k}}^{(1)}$ reads
\begin{eqnarray}
 f_{\pm\mathbf{k}}^{(1)} & = &  
 \pm \hbar v k_x \left(\frac{\mu{\cal E}_x}{v}\right) \left(-\frac{d f_{\pm \mathbf{k}}^{(0)}}{dE_{\pm k}}\right).
 \label{f1}
\end{eqnarray}
Here, we assume long-range electron scattering in graphene \cite{trushin2008conductivity}, which results in the momentum-independent mobility $\mu$
and drift velocity $\mu {\cal E}_x$ \cite{SM}.
The electric field ${\cal E}_x$ applied along the x-axis swipes the conduction-band electrons 
and the valence-band holes to positive $k_x$ creating inverse population, as depicted in Fig. \ref{fig1}(d).
At the same time, the electrons and holes gain an average momentum determined by the extrema of $f_{\pm\mathbf{k}}^{(1)}$,
which occur at $k=\pi k_B T_e/(2\hbar v)$ with $k_B$ being the Boltzmann constant. This value determines the lengths of the vectors $\mathbf{k}$ and $\mathbf{k}'$
in Fig. \ref{fig1}(d).
The SPP occupation can be written in a similar way as $N_\mathbf{q}=N_\mathbf{q}^{(0)}+N_\mathbf{q}^{(1)}$,
where $N_\mathbf{q}^{(0)}$ is the Bose-Einstein distribution at the temperature $T_0$,
and $N_\mathbf{q}^{(1)}$ is the non-equilibrium term that can be found from the rate equation given by
\begin{eqnarray}
 \frac{d N_\mathbf{q}}{d t}& = & \sum\limits_{\mathbf{k}} \left(W^\mathrm{emi}_{\mathbf{k}\to \mathbf{k}-\mathbf{q}}
 -  W^\mathrm{abs}_{\mathbf{k}\to \mathbf{k}+\mathbf{q}}\right)\\
 & = & G_\mathbf{q} - \frac{N_\mathbf{q}^{(1)}}{\tau_\mathbf{q}},
 \label{rate}
\end{eqnarray}
where $W^\mathrm{emi(abs)}_{\mathbf{k}\to \mathbf{k}\mp\mathbf{q}}$ is the golden-rule emission (absorption) probability
calculated from Eqs. (\ref{transit}), $G_\mathbf{q}$ is the SPP generation rate, and 
$\tau_\mathbf{q}$ is the extrinsic SPP decay time due to the absorption by electrons in graphene, as shown in Fig. \ref{fig2}.
The steady-state solution of Eq. (\ref{rate}), which explicitly incorporates the intrinsic decay rate, along with its justification,
is presented in the End Matter and plotted in Fig. \ref{fig3}(a,b) in comparison with $N_\mathbf{q}^{(0)}$.

\begin{figure}
 \includegraphics[width=\columnwidth]{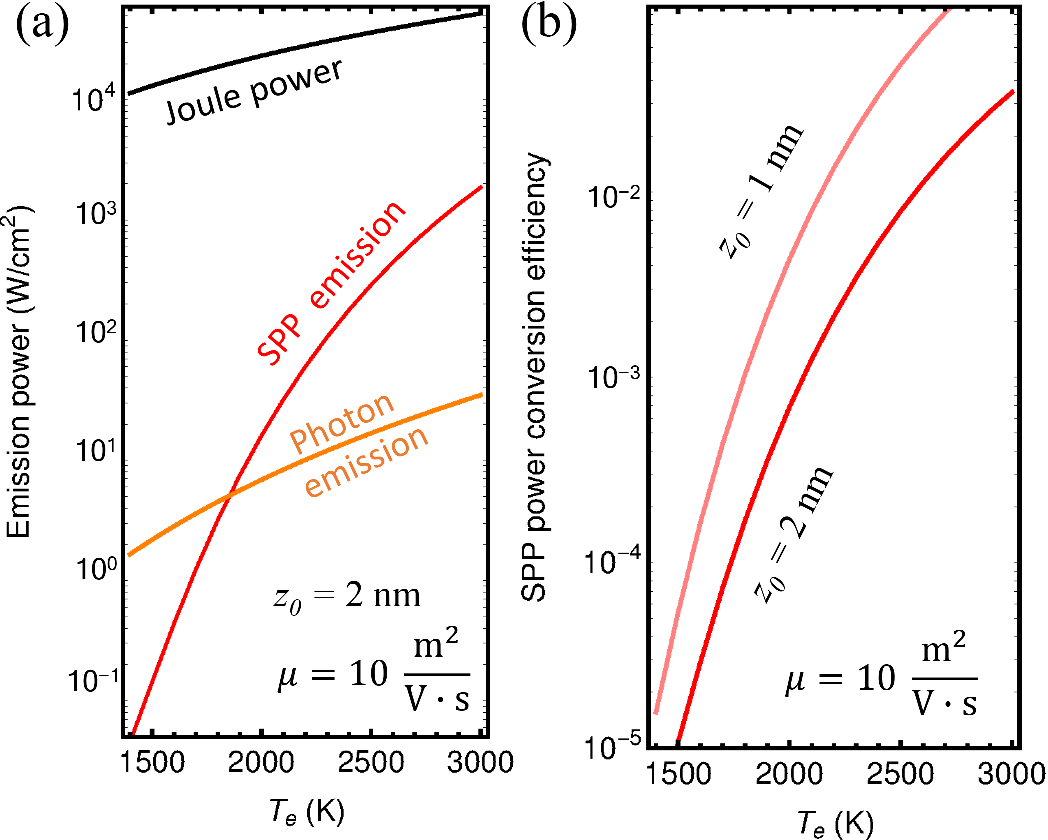}
 \caption{
 SPP generation efficiency in the high-bias regime ($\mu {\cal E}_x/v = 0.9$).
 (a) SPP emission power (red curve) compared with photon emission (orange curve) and electrical (Joule) power (black curve). Note the exponential increase in SPP emission between 1500 and 2000 K, surpassing photon emission at around 1800 K.
 (b) SPP power conversion efficiency computed for different separations between the hBN/Ag interface and graphene is comparable or higher than
  the typical literature values \cite{zhang2021electrical,pommier2023nanoscale,wang2023engineering}.
 \label{fig4}
 }
\end{figure}

{\it Discussion. ---} Figure \ref{fig3} demonstrates the effect of strong bias and high electron temperature on SPP emission.
The emission peak is determined by the total electron and hole energy, which is released upon recombination.
The electron energy can be estimated as a sum of the thermal energy, $2k_BT_e$, and the energy gained 
by the same electrons between two subsequent scattering events, $e{\cal E}_x l_\mathrm{mfp}$. 
The electron mean free path can be related to the electron mobility as $l_\mathrm{mfp}=\hbar k \mu/e$, where
electron momentum is given by $\hbar k\sim 2k_BT_e/v$. The same is true for holes, and 
the total electron-hole recombination energy released in peak SPP emission reads 
\begin{equation}
 \hbar\omega_{spp}^\mathrm{peak} \sim 4k_B T_e\left(1+\frac{\mu {\cal E}_x}{v}\right).
 \label{spp-max}
\end{equation}
In the conventional conductors, the drift velocity is always small ($\mu {\cal E}_x\ll v$) and electron temperature is low ($T_e\sim T_0$).
As a consequence, the electron-hole recombination energy is always within the thermal smearing of the SPP bath and therefore
hard (if not impossible) to detect. Since graphene can sustain very high bias, electron temperature can reach 2800K \cite{2800K-2023suspend,2800K-2015suspend}
in suspended samples and the drift velocity can approach the band velocity \cite{drift2011theory,drift2017hBN}.
Equation (\ref{spp-max}) suggests and Fig. \ref{fig3}(a) confirms that the SPP emission spectral maximum 
approaches 0.6 eV at $\mu {\cal E}_x \to v$ and $T_e$ below 1000K. 
Figure \ref{fig3}(b) illustrates further increase of $\omega_{spp}^\mathrm{peak}$ with $T_e$.
Adjusting bias voltage we can theoretically cover the whole near-infrared SPP emission spectral range (0.41--1.58 eV).
Most importantly, the higher $T_e$ increases the emission intensity by orders of magnitude, cf. Fig. \ref{fig3}(a) and (b).

To further characterize the SPP emission we compute the total emission rate, $\Gamma_{spp}$, and power, $P_{spp}$, given respectively by 
 \begin{eqnarray}
 && \left\{
\begin{array}{c}
\Gamma_{spp} \\
P_{spp}
\end{array}
\right\}
 = \int\limits_0^{2\pi}d\theta\int\limits_0^{\omega_{sp}}\frac{d\omega q}{(2\pi)^2} \frac{dq}{d\omega}
 \left\{
\begin{array}{c}
G_\mathbf{q} \\
\hbar \omega G_\mathbf{q}
\end{array}
\right\},
 \label{Gamma}
\end{eqnarray}
where $\tan\theta=q_y/q_x$.
Both quantities are normalized to the unit interfacial area. SPP and photon emission are two competing processes, and their relative strength is quantified by the Purcell factor $F=\Gamma_{spp}/\Gamma_{ph}$ \cite{qian2021spontaneous,iwase2010analysis}, where 
$\Gamma_{ph}$ is the total photon emission into free space, as discussed in standard texts \cite{vasko1998electronic} and outlined in Ref. \cite{SM}.
Importantly, the SPP density of states diverges as the emission energy approaches the surface plasmon energy quantum $\hbar\omega_{sp}$, 
see Fig. \ref{fig1}(c), while this divergence does not occur for free photons.
As a result, the Purcell factor increases dramatically when higher temperatures cause the high-energy tail of the emission spectrum to overlap with the threshold energy 
$\hbar\omega_{sp}$, see Fig. \ref{fig3}(b). The Purcell factor cannot be enhanced simply by increasing the number of electron-hole pairs
because this would also increase photon emission \cite{SM}.


The Purcell factor is the ``optical'' quantity and does not describe the power conversion efficiency of the device. The latter can be written as
$f=P_{spp}/P_{tot}$, where $P_{tot}=P_{el}+P_{ph}+P_{spp}$. Here, $P_{spp}$ is given by Eq. (\ref{Gamma}), 
$P_{ph}$ is the photon emission power derived in Ref. \cite{SM}, and $P_{el}={\cal E}_x j_x$ is the electrical (Joule) power
with $j_x$ being the electrical current density in graphene. Using Eq. (\ref{f1}) we can deduce $j_x$ easily, and $P_{el}$ reads
\begin{equation}
 P_{el} = \frac{e\pi k_B^2 T_e^2}{12\hbar^2 \mu}\left(\frac{\mu{\cal E}_x}{v}\right)^2.
\end{equation}
$P_{el}$, $P_{ph}$, and $P_{spp}$ must be considered consistently in the same regime $\mu {\cal E}_x/v\to 1$ (Fig. \ref{fig4}).
The Joule power dominates the photon and SPP emission but it is a relatively weak (quadratic) function of $T_e$, whereas $P_{ph}$ and $P_{spp}$ are respectively quartic and exponential functions of $T_e$. Thus, $P_{spp}$ overrides $P_{ph}$ above a certain $T_e$ and turns out to be only about two orders magnitude lower than $P_{el}$
in the extremely high-temperature limit. This limits the power conversion efficiency to about 1\%, which is an excellent figure for SPP generation efficiency \cite{zhang2021electrical,pommier2023nanoscale,wang2023engineering}.

{\it Conclusion. ---} 
The key novelty of the proposed concept lies in the spatial separation of the SPP emitter from the SPP-supporting interface that
is only achievable because both the electrons in graphene and the SPPs themselves are essentially 2D, each confined and propagating within its respective plane.
While hyperbolic phonon polaritons can also be generated in hBN using strongly biased graphene \cite{abou2025electroluminescence,guo2025hyperbolic},
their inherently bulk nature precludes spatial separation between the emitting and hosting regions complicating optoelectronic applications.
Theoretically, nanoantennas can {\em locally} emit SPPs with even higher Purcell factor and conversion efficiency \cite{zhang2019theory}, while
the proposed setup offers a {\em uniform} emission giving a unique opportunity to study SPP diffraction, interference, and gyrotropic phenomena directly at the interface
 \cite{Review2010,APL2002SPPoptics,APL2007diffractive,PRL2005far-field,kolesov2009wave,katsantonis2023giant}.
 However, it should be noted that the near-field tips typically used to detect SPPs \cite{hillenbrand2025visible}, would locally disrupt this uniformity,
 see a focused feasibility discussion in the End Matter.

\acknowledgments

This research is supported by the Singapore Ministry of Education Research Centre of Excellence award to the Institute for Functional Intelligent Materials
(I-FIM, Project No. EDUNC-33-18-279-V12).
I thank Zhe Wang, Vijith Kalathingal, and Goki Eda for discussions.

\bibliography{SPP.bib,parameters.bib}

\appendix

\section{End Matter}
\setcounter{equation}{0}
\renewcommand{\theequation}{A\arabic{equation}}

{\em Hamiltonian. ---} The Hamiltonian describing electrical SPP generation comprises the SPP, electronic, and electron-SPP interaction terms given by \cite{archambault2010quantum,bernardi2015theory}
\begin{eqnarray}
   H & = & H_{spp} + H_e + H_{e-spp}, \\
  H_{spp} & = & \sum\limits_{\mathbf{q}}\hbar\omega\left(a_\mathbf{q}^\dagger a_\mathbf{q} + 1/2 \right),\\
H_e & = & \sum\limits_{\mathbf{k}}\left(\varepsilon_k c_{+\mathbf{k}}^\dagger c_{+\mathbf{k}} -
 \varepsilon_k c_{-\mathbf{k}}^\dagger c_{-\mathbf{k}}\right),\\
\nonumber H_{e-spp} & = & -\sqrt{\frac{\pi e^2 \hbar}{2\omega L_x L_y L_z}}e^{-z_0/l_d}\\
&& \times
 \sum\limits_{\mathbf{k},\mathbf{q}}\mathbf{e}_\mathbf{q}\cdot \mathbf{v}(\mathbf{k},\mathbf{q})
 \left(a_\mathbf{q}+ a_\mathbf{q}^\dagger \right).
 \label{HSPP}
\end{eqnarray}
Here, $a_\mathbf{q}^\dagger$ and $a_\mathbf{q}$ denote the SPP creation and annihilation operators.
The operators $c_{+\mathbf{k}}^\dagger$ and $c_{-\mathbf{k}}^\dagger$ create an electron 
with the wave vector $\mathbf{k}$ in conduction and valence bands, respectively, while
$c_{\pm\mathbf{k}}$ annihilate an electron in the corresponding bands.
The electron dispersion is $\varepsilon_k=hvk$.
The factor $e^{-z_0/l_d}$ accounts for the separation between graphene and the plasmonic interface.
The quantities $L_{x,y,z}$ are the normalization lengths,
$e$ is the elementary charge, $\mathbf{e}_\mathbf{q}$ is the unit vector along the SPP propagation direction,
and $\mathbf{v}(\mathbf{k},\mathbf{q})$ is the electron velocity operator with the $x$ and $y$ components given by
\begin{equation}
 v_{x,y}(\mathbf{k},\mathbf{q})=v\left( c_{+(\mathbf{k}+\mathbf{q})}^\dagger, c_{-(\mathbf{k}+\mathbf{q})}^\dagger \right)
M_{x,y}^{\phi\phi'} \left(
\begin{array}{c}
c_{+\mathbf{k}} \\
c_{-\mathbf{k}}
\end{array}\right),
\label{matrixV}
\end{equation}
where
\begin{equation}
\nonumber M_x^{\phi\phi'}=   \frac{1}{2}\left(
\begin{array}{cc}
e^{i\phi}+e^{-i\phi'} & -e^{i\phi}+e^{-i\phi'} \\
e^{i\phi}-e^{-i\phi'} &  -e^{i\phi}-e^{-i\phi'}
   \end{array}
   \right),
\end{equation}
\begin{equation}
\nonumber M_y^{\phi\phi'}=  \frac{1}{2i} \left(
\begin{array}{cc}
e^{i\phi}-e^{-i\phi'} & -e^{i\phi}-e^{-i\phi'} \\
e^{i\phi}+e^{-i\phi'} &  -e^{i\phi}+e^{-i\phi'}
   \end{array}
   \right),
\end{equation}
$\tan\phi=k_y/k_x$, $\tan\phi'=(k_y+q_y)/(k_x+q_x)$. 
The matrices $M_{x,y}^{\phi\phi'}$ arise from the chiral structure inherited by the electronic states 
from the honeycomb lattice symmetry of graphene. Only off-diagonal terms contribute to the interband transitions considered here.

While $L_x$ and $L_y$ are set by the in-plane geometry of the interface, $L_z$ depends on the field penetration lengths $l_{m,d}$ \cite{archambault2010quantum,SM} and reads
\begin{equation}
\label{Lz}
 L_z = \frac{l_d}{2}\left(\frac{\omega_p^2}{\omega^2} -\epsilon_\infty \right)
+\frac{l_m}{2}\left(\frac{\omega_p^2}{\omega^2} 
+ \frac{\epsilon_d \epsilon_\infty}{\frac{\omega_p^2}{\omega^2} -\epsilon_\infty}\right),
\end{equation}
where
\begin{eqnarray}
 l_d & = & \frac{c}{\omega \epsilon_d} \sqrt{\frac{\omega_p^2}{\omega^2}- (\epsilon_\infty+\epsilon_d)},\\
 l_m & = & \frac{c}{\omega}\frac{\sqrt{\frac{\omega_p^2}{\omega^2}- (\epsilon_\infty+\epsilon_d)}}{\frac{\omega_p^2}{\omega^2}-\epsilon_\infty}.
\end{eqnarray}
As a function of $\omega$, $L_z$ approximately follows  $l_d$.

{\em Balance of rates. ---} These ingredients enable calculation of the golden-rule SPP generation and extrinsic decay rates.
In the steady-state regime, the left-hand side of Eq. (\ref{rate}) equals to $1/\tau_0$, where $\tau_0$ is the intrinsic 
SPP decay time of several tens of fs due to electrons in the metal underneath the interface.
The full expressions for $\tau_\mathbf{q}$ and $G_\mathbf{q}$ can be found in Ref. \cite{SM}. The approximate expressions read
\begin{eqnarray}
 \label{tauq} \frac{1}{\tau_\mathbf{q}} & = &\frac{\pi e^2 \omega\, \mathrm{e}^{-2 z_0/l_d}}{16\hbar L_z \sqrt{\omega^2-q^2v^2}},\\
\nonumber G_\mathbf{q} &= &  \frac{e^2  e^{-2 z_0/l_d}}{16 \hbar\omega L_z}\int\limits_0^{2\pi}d\phi 
\frac{\omega^2 (\omega^2-q^2 v^2)\sin^2(\theta-\phi)}{\left[\omega - q v \cos(\theta-\phi)\right]^3} \\
 &&  \times \left[-f_{+\mathbf{k}}^{(1)}f_{-(\mathbf{k}-\mathbf{q})}^{(1)} \right]_{k=\frac{1}{2v} \frac{\omega^2 - q^2 v^2 }{\omega- q v \cos(\theta-\phi)}},
 \label{Gq}
\end{eqnarray}
where $\tan\theta=q_y/q_x$, $\tan\phi=k_y/k_x$.
Note that $G_\mathbf{q}$ is quadratic in ${\cal E}_x$ because electron-hole recombination requires both an excited electron state and an emptied hole state produced by the same electric field.

The extrinsic decay rate increases with $\omega$, reaches a maximum of the order of $\tau_0$,
and then drops abruptly when $\omega \to\omega_{sp}$ because SPPs decouple from graphene ($l_d\to 0$), see Fig. \ref{fig2}(a,b). 
Consequently, the interfacial (intrinsic) life-time $\tau_0\lesssim 10^{-13}$s dominates 
the total SPP relaxation time $\tau=\tau_0\tau_\mathbf{q}/(\tau_0+\tau_\mathbf{q})$ within the SPP emission spectrum. 
In contrast, the non-equilibrium electron-hole distribution (\ref{f1}) relaxes at the time scale of
$(2k_B T_e\mu)/(ev^2)\sim 10^{-12}$s for $\mu\sim10^5$ cm$^2$/(V$\cdot$s) and $T_e\sim 1000$K.
Therefore, during the typical SPP lifetime $\tau_0$, the electron-hole distribution is effectively constant, ensuring continuous SPP emission.
This justifies the steady-state treatment, and solution of Eq. (\ref{rate}) is simply given by $N_\mathbf{q}^{(1)} = \tau G_\mathbf{q}$.

{\em Feasibility. ---} The core device is a standard hBN/graphene/hBN heterostructure \cite{above2000K-2017hBN} capped with an ultrathin gold film.
The gold layer should be only a few nanometers thick to permit SPP detection through its thickness. Such films are routinely fabricated by XPANCEO, and transfer onto glass substrates has been demonstrated \cite{mironov2024graphene}. These semitransparent films retain the essential low-loss plasmonic properties required here \cite{yakubovsky2023optical}. 
Operational electron temperatures of about 2000 K can be achieved in hBN/graphene/hBN devices on quartz under 10 V bias in ambient conditions \cite{above2000K-2017hBN}. 
Strongly out-of-equilibrium carrier distributions with drift velocity approaching the band velocity are reached in undoped graphene on hBN under comparable biases \cite{drift2017hBN}. The hBN spacer between graphene and gold should be 2-3 monoloayers ($\sim$1--2 nm) to enable efficient coupling between electron-hole dipoles
in graphene and SPPs at the Au/hBN interface while maintaining electrical insulation \cite{illarionov2020insulators}.
For detection, s-SNOM may be employed \cite{hillenbrand2025visible}, using the AFM tip solely for SPP detection (not excitation) to minimize long-range signals from the gold/air interface. As in recent Au/MoS$_2$ studies \cite{yakubovsky2023optical}, SPP emission can also be probed at cracks and edges.

\section{Supplemental material}
\setcounter{equation}{0}
\renewcommand{\theequation}{S\arabic{equation}}
\setcounter{figure}{0}
\renewcommand{\thefigure}{S\arabic{figure}}

\section{S1. SPP basics}

According to the Maxwell's equations, SPPs can propagate along the metal-dielectric interface
assuming that the metallic dielectric function $\epsilon_m$ acquires negative values at low frequencies, as described by
\begin{equation}
 \epsilon_m (\omega) = \epsilon_\infty - \frac{\omega_p^2}{\omega^2},
 \label{p}
\end{equation}
where $\epsilon_\infty$ is the high-frequency dielectric permittivity.
The dielectric permittivity of an insulator is always positive, $\epsilon_d>0$.
To find the electric field at the interface we have
to satisfy the Maxwell's equations and match continuity at $z=0$.
The result reads \cite{zayats2005nano}
\begin{eqnarray}
 \mathbf{B} &= & \left\{
\begin{array}{ll}
\nonumber (0,B_0,0){\mathrm e}^{iqx-\kappa_d z -i\omega t}, & z>0,\\
\nonumber  (0, B_0, 0){\mathrm e}^{iqx+\kappa_m z -i\omega t}, & z<0;
\end{array}
 \right.\\
\mathbf{E} & = & \left\{
\begin{array}{ll}
\nonumber  \left(-\frac{B_0 c  \kappa_d}{i\omega \epsilon_d},0,-\frac{B_0 c q}{\omega \epsilon_d}\right){\mathrm e}^{iqx-\kappa_d z -i\omega t}, & z>0,\\
\nonumber  \left(\frac{B_0 c  \kappa_m}{i\omega \epsilon_m}, 0, -\frac{B_0 c q}{\omega \epsilon_m}\right){\mathrm e}^{iqx+\kappa_m z -i\omega t}, & z<0;
\end{array}
 \right.\\
 \label{field}
\end{eqnarray}
\begin{eqnarray}
\label{s}
 q^2 - \kappa_d^2 & = & \frac{\omega^2}{c^2}\epsilon_d,\\
 q^2 - \kappa_m^2 & = & \frac{\omega^2}{c^2}\epsilon_m,
 \label{m}
\end{eqnarray}
and
\begin{equation}
 \frac{\kappa_m}{\kappa_d} = - \frac{\epsilon_m}{\epsilon_d}.
 \label{main}
\end{equation}
Here, $l_d=1/\kappa_d$ and $l_m=1/\kappa_m$ are the penetration length of
the field into the dielectric and metal, respectively.
Obviously, to satisfy equation (\ref{main}) the dielectric function 
$\epsilon_m$ must be negative so that $\omega<\omega_p$.
Combining equations (\ref{p}), (\ref{s}), (\ref{m}), and (\ref{main}) we find the
SPP dispersion relation given by {\bf Eq. (1)} in the main text.
To keep SPPs propagating the frequency must be 
$\omega<\omega_{sp}$, where $\omega_{sp}=\omega_p/\sqrt{\epsilon_\infty+\epsilon_d}$
is the surface plasmon frequency.
In the low-frequency limit, $\omega\to 0$, we have  $q\to \sqrt{\epsilon_d}\omega/c$,
which is the photon dispersion in the dielectric media at $z>0$.

It is also instructive to write $l_d$ and $l_m$ explicitly as
\begin{eqnarray}
 l_d & = & \frac{c}{\omega \epsilon_d} \sqrt{\frac{\omega_p^2}{\omega^2}- (\epsilon_\infty+\epsilon_d)},\\
 l_m & = & \frac{c}{\omega}\frac{\sqrt{\frac{\omega_p^2}{\omega^2}- (\epsilon_\infty+\epsilon_d)}}{\frac{\omega_p^2}{\omega^2}-\epsilon_\infty}.
 \end{eqnarray}
Note that $l_{d,m}\to 0$ at $\omega\to \omega_{sp}$, and $l_m \ll l_d$ at $\omega\to 0$.
For our purpose, $l_d$ must be substantially larger than $z_0$.
The lengths $l_{m,d}$ are depicted in {\bf Fig. 1(b)} (main text).

The SPP wave has two components: the in-plane ($E_x$) and out-of-plane ($E_z$) ones.
The out-of-plane component is normal to the wave vector (like in the case of conventional electromagnetic waves),
whereas the in-plane component is longitudinal (along the SPP wave vector $\mathbf{q}$).
At $z>0$, the ratio between in-plane and out-of-plane components is given by
\begin{eqnarray}
\nonumber
 \frac{|E_x|}{|E_z|}& = & \frac{1}{l_d q}\\
 \nonumber
 & = & \sqrt{\frac{\epsilon_d}{\frac{\omega_p^2}{\omega^2}-\epsilon_\infty}}\\
 & = & \sqrt{\frac{l_m}{l_d}}.
\label{relation}
\end{eqnarray}
Hence, $|E_x|\ll |E_z|$ at low frequency, but $|E_x|\to |E_z|$ at $\omega\to \omega_{sp}$.
The later limit makes it possible to couple the SPP field with in-plane electron motion in graphene.
At $z<0$, the ratio is inverted and given by
\begin{equation}
 \frac{|E_x|}{|E_z|} = \sqrt{\frac{l_d}{l_m}}.
\end{equation}

\begin{figure}
 \includegraphics[width=0.8\columnwidth]{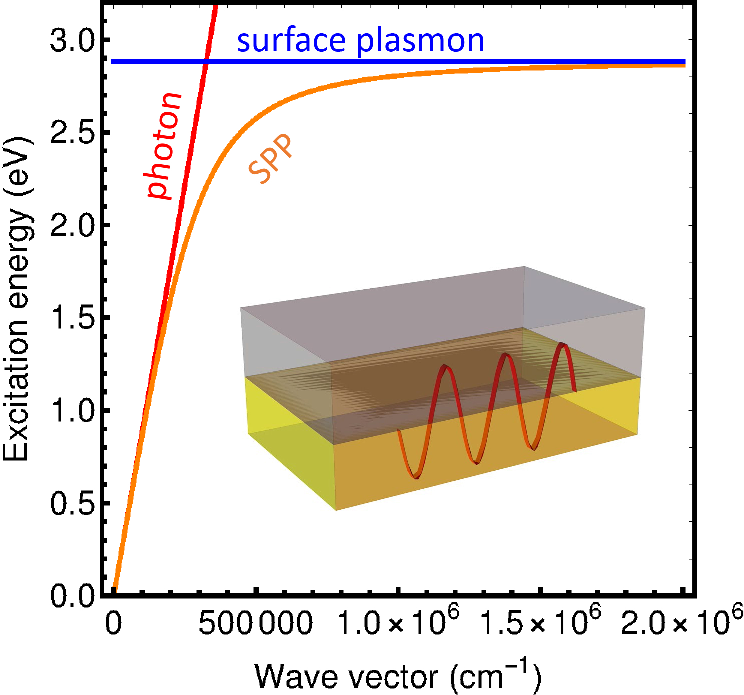}
 \caption{SPP dispersion contrasted to the photon and surface plasmon dispersions for the Ag/hBN interface.  }
\end{figure}

\section{S2. SPP quantization}

To describe SPP generation we have to quantize the SPP field. 
The SPP Hamiltonian can be written as
\begin{equation}
 H_{SPP} = \sum\limits_{\mathbf{q}}\hbar\omega\left(a_\mathbf{q}^\dagger a_\mathbf{q} + 1/2 \right),
\end{equation}
where $a_\mathbf{q}^\dagger$ and $a_\mathbf{q}$ are the SPP creation and annihilation operators,
and $\omega$ can be found from the SPP dispersional relation $q(\omega)$ for each mode $\mathbf{q}$.
We introduce the mode occupation number, $N_\mathbf{q}$, which occurs in the operator rules as 
\begin{eqnarray}
 a_\mathbf{q}^\dagger |N_\mathbf{q}\rangle & = & \sqrt{N_\mathbf{q}+1}|N_\mathbf{q}+1\rangle,\\
 a_\mathbf{q} |N_\mathbf{q}\rangle & = & \sqrt{N_\mathbf{q}}|N_\mathbf{q}-1\rangle.
\end{eqnarray}
The creation and annihilation operators are related to the vector potential as \cite{archambault2010quantum}
\begin{equation}
A_\mathbf{q}  \leftrightarrow a_\mathbf{q}\sqrt{\frac{2\pi \hbar c^2}{\omega L_x L_y}},\quad
A_\mathbf{q}^*  \leftrightarrow a_\mathbf{q}^\dagger \sqrt{\frac{2\pi \hbar c^2}{\omega L_x L_y}},
\end{equation}
where $L_x$ and $L_y$ are the sample length and width.
We use the Gauss units so that
$\mathbf{E}$ and $\mathbf{B}$ fields have the same dimensionality.
The out-of-plane renormalization length, $L_z$, 
can be found from the SPP energy density normalization condition given by \cite{archambault2010quantum}
\begin{equation}
 \frac{1}{2}\int\limits_{-\infty}^{\infty}dz \left[\frac{d(\omega \epsilon)}{d\omega}
 |\mathbf{E}_\mathbf{q}|^2 + |\mathbf{B}_\mathbf{q}|^2\right]
 = \frac{\omega^2}{c^2} |\mathbf{A}_\mathbf{q}|^2,
 \label{normLz}
\end{equation}
where
\begin{equation}
 \epsilon=\left\{
\begin{array}{ll}
\epsilon_d, & z>0,\\
\epsilon_m(\omega), & z<0,
\end{array}
 \right.
\end{equation}
and 
\begin{eqnarray}
 \mathbf{B}_\mathbf{q} &= & \frac{i\omega A_\mathbf{q}}{c\sqrt{L_z}}  \left\{
\begin{array}{ll}
\nonumber  (0,-\frac{i\omega \epsilon_d}{c\kappa_d},0){\mathrm e}^{-\kappa_d z}, & z>0,\\
\nonumber  (0, \frac{i\omega \epsilon_m}{c\kappa_m}, 0){\mathrm e}^{\kappa_m z}, & z<0;
\end{array}
 \right.\\
\mathbf{E}_\mathbf{q} & = & \frac{i\omega A_\mathbf{q}}{c\sqrt{L_z}} \left\{
\begin{array}{ll}
\nonumber  \left(1,0,\frac{i q}{\kappa_d}\right){\mathrm e}^{-\kappa_d z}, & z>0, \\
\nonumber  \left(1, 0, -\frac{i q}{\kappa_m}\right){\mathrm e}^{\kappa_m z}, & z<0;
\end{array}
 \right.\\
 \label{fieldQ}
\end{eqnarray}
with 
\begin{eqnarray}
\label{Er}
\mathbf{E}(\mathbf{r},t) & = &  \sum\limits_{\mathbf{q}} \mathbf{E}_\mathbf{q}{\mathrm e}^{iqx -i\omega t},\\
\label{Br}
\mathbf{B}(\mathbf{r},t) & = &  \sum\limits_{\mathbf{q}} \mathbf{B}_\mathbf{q}{\mathrm e}^{iqx -i\omega t},\\
\label{Dr}
\mathbf{D}(\mathbf{r},t) & = & \epsilon \mathbf{E}(\mathbf{r},t),\\
\label{Hr} 
\mathbf{H}(\mathbf{r},t) & = & \mathbf{B}(\mathbf{r},t).
\end{eqnarray}
Equations (\ref{Er}--\ref{Hr}) satisfy the Maxwell's equations 
\begin{eqnarray}
\nabla \times \mathbf{E}(\mathbf{r},t) & = &  -\frac{1}{c} \frac{\partial \mathbf{B}(\mathbf{r},t)}{\partial t},\\
\nabla\times \mathbf{B}(\mathbf{r},t) & = & \frac{1}{c} \frac{\partial \mathbf{D}(\mathbf{r},t)}{\partial t} ,\\
\nabla\cdot \mathbf{D}(\mathbf{r},t) & = & 0,\\
\nabla\cdot \mathbf{B}(\mathbf{r},t) & = & 0.
\end{eqnarray}
as long as conditions (\ref{s}), (\ref{m}),  and (\ref{main}) are fulfilled.

The squared fields reads
\begin{eqnarray}
 |\mathbf{B}_\mathbf{q}|^2 &= & \frac{\omega^2 |A_\mathbf{q}|^2}{c^2 L_z}  \left\{
\begin{array}{ll}
\nonumber  \frac{\omega^2 \epsilon_d^2}{c^2 \kappa_d^2}{\mathrm e}^{-2\kappa_d z}, & z>0,\\
\nonumber  \frac{\omega^2 \epsilon_m^2}{c^2 \kappa_m^2}{\mathrm e}^{2\kappa_m z}, & z<0;
\end{array}
 \right.\\
|\mathbf{E}_\mathbf{q}|^2 & = & \frac{\omega^2 |A_\mathbf{q}|^2}{c^2 L_z} \left\{
\begin{array}{ll}
\nonumber \left(1+\frac{q^2}{\kappa_d^2}\right){\mathrm e}^{-2\kappa_d z}, & z>0, \\
\nonumber  \left(1+\frac{q^2}{\kappa_m^2}\right){\mathrm e}^{2\kappa_m z}, & z<0;
\end{array}
 \right.\\
 \label{fieldQ2}
\end{eqnarray}
and the normalization condition (\ref{normLz}) takes the form \cite{archambault2010quantum}
\begin{eqnarray}
\nonumber L_z &=& \frac{1}{4\kappa_d}\left[\epsilon_d\left(1+\frac{q^2}{\kappa_d^2}\right)+ \frac{\omega^2 \epsilon_d^2}{c^2\kappa_d^2}\right]\\
&& +\frac{1}{4\kappa_m}\left[\left(1+\frac{q^2}{\kappa_m^2}\right)\frac{d(\omega \epsilon_m)}{d\omega}
+ \frac{\omega^2 \epsilon_m^2}{c^2\kappa_m^2}\right].
\label{Lz1}
\end{eqnarray}
Note that $\epsilon_d=1$ in Ref. \cite{archambault2010quantum} that makes the normalization length slightly different from equation (\ref{Lz1}).
It is instructive to rewrite $L_z$ explicitly in terms of $l_d$ and $l_m$, see {\bf Eq. (A6)} in the main text.


\section{S3. Electron--SPP interaction}

To describe interactions between SPPs and electrons in graphene, we 
first introduce the low-energy tight-binding Hamiltonian for electrons given by
\begin{equation}
 \hat H_0=\left(
\begin{array}{cc}
0 & \hbar v (\hat k_x-i \hat k_y)\\
\hbar v (\hat k_x+i\hat k_y) &  0
   \end{array}
   \right).
   \label{Hparallel}
\end{equation}
Here, $v\sim 10^8$ cm/s is a constant, and $\hbar \hat k_{x,y}$ are electron momentum operators.
Hamiltonian (\ref{Hparallel}) describes a single valley/spin channel. 
To consider interactions between SPPs and electrons we need the electron velocity operator given by
\begin{equation}
 \hat v=\mathbf{e}_x\left(
\begin{array}{cc}
0 & v\\
v  &  0
   \end{array}
   \right)
   +\mathbf{e}_y \left(
\begin{array}{cc}
0 & -iv \\
+i v &  0
   \end{array}
   \right),
\end{equation}
where  $\mathbf{e}_{x,y}$ are the unit vectors.
The eigenfunctions of $\hat H_0$ can be written explicitly as 
$\chi_\pm(x,y)$, where  ``$\pm$'' refers to the conduction and valence bands, respectively, and
\begin{equation}
\chi_\pm(x,y)=\frac{1}{\sqrt{2L_xL_y}}\mathrm{e}^{ik_x x+ik_y y}
\left(
\begin{array}{c}
1\\
\pm \mathrm{e}^{i\phi}
\end{array}\right),
\label{chi+}
\end{equation}
where  $\tan\phi=k_y/k_x$, $k=\sqrt{k_x^2+k_y^2}$.
The corresponding eigenvalues are given by $E_{\pm k}=\pm\varepsilon_k$, where $\varepsilon_k=\hbar v k$.
The wave functions are normalized as
\begin{eqnarray}
 \int\limits_0^{L_x} dx \int\limits_0^{L_y} dy \chi_\pm^*(x,y)\chi_\pm(x,y)=1.
 \label{norm}
\end{eqnarray}
Now, we can write the diagonalized Hamiltonian in the second quantization form as
\begin{equation}
 H_0= \sum\limits_{\mathbf{k}}\left(\varepsilon_k c_{+\mathbf{k}}^\dagger c_{+\mathbf{k}} -
 \varepsilon_k c_{-\mathbf{k}}^\dagger c_{-\mathbf{k}}\right),
\end{equation}
where $c_{+\mathbf{k}}^\dagger$ ($c_{-\mathbf{k}}^\dagger$) creates an electron 
with the wave vector $\mathbf{k}$ in conduction (valence) band,
and $c_{\pm\mathbf{k}}$ removes an electron in the respective bands.

The electron-SPP interaction Hamiltonian can be written in terms of the SPP creation/annihilation operators,
($a_\mathbf{q}^\dagger$, $a_\mathbf{q}$), and velocity operator $\mathbf{v}(\mathbf{k},\mathbf{q})$ as
\begin{equation}
 H_1=-\frac{e}{2c}\sqrt{\frac{2\pi \hbar c^2}{\omega L_x L_y L_z}}e^{-\kappa_dz_0}
 \sum\limits_{\mathbf{k},\mathbf{q}}\mathbf{e}_\mathbf{q}\cdot \mathbf{v}(\mathbf{k},\mathbf{q})
 \left(a_\mathbf{q}+ a_\mathbf{q}^\dagger \right).
 \label{H1}
\end{equation}
Here, $e^{-\kappa_dz_0}$ takes into account graphene separation from the dielectric-metal interface, and 
$\mathbf{e}_\mathbf{q}$ is the unit vector indicating the SPP propagation direction, which coincides with the in-plane polarization vector.
The scalar product $\mathbf{e}_\mathbf{q}\cdot \mathbf{v}(\mathbf{k},\mathbf{q})$ can be explicitly written as
\begin{equation}
\mathbf{e}_\mathbf{q}\cdot \mathbf{v}(\mathbf{k},\mathbf{q}) = 
v_x(\mathbf{k},\mathbf{q})\,\cos\theta  + v_y(\mathbf{k},\mathbf{q})\,\sin\theta,
\end{equation}
where $\tan\theta=q_y/q_x$.

The electron velocity operator can also be written in terms of the electron creation/annihilation operators as
\begin{equation}
 v_{x,y}(\mathbf{k},\mathbf{q})=v\left( c_{+(\mathbf{k}+\mathbf{q})}^\dagger, c_{-(\mathbf{k}+\mathbf{q})}^\dagger \right)
M_{x,y}^{\phi\phi'} \left(
\begin{array}{c}
c_{+\mathbf{k}} \\
c_{-\mathbf{k}}
\end{array}\right),
\end{equation}
where
\begin{equation}
M_x^{\phi\phi'}=   \left(
\begin{array}{cc}
\frac{e^{i\phi}+e^{-i\phi'}}{2} & \frac{-e^{i\phi}+e^{-i\phi'}}{2} \\
\frac{e^{i\phi}-e^{-i\phi'}}{2} &  \frac{-e^{i\phi}-e^{-i\phi'}}{2}
   \end{array}
   \right),
\end{equation}
\begin{equation}
M_y^{\phi\phi'}=   \left(
\begin{array}{cc}
\frac{e^{i\phi}-e^{-i\phi'}}{2i} & \frac{-e^{i\phi}-e^{-i\phi'}}{2i} \\
\frac{e^{i\phi}+e^{-i\phi'}}{2i} &  \frac{-e^{i\phi}+e^{-i\phi'}}{2i}
   \end{array}
   \right),
\end{equation}
and  $\tan\phi'=(k_y+q_y)/(k_x+q_x)$.

We consider interband transitions only so that in the case of SPP emission
the initial state reads $|n_{-(\mathbf{k}-\mathbf{q})},n_{+\mathbf{k}};N_\mathbf{q}\rangle$,
and the final state is given by 
$|n_{-(\mathbf{k}-\mathbf{q})}+1,n_{+\mathbf{k}}-1;N_\mathbf{q}+1 \rangle$.
In the case of  SPP absorption
the initial and final states read $|n_{+(\mathbf{k}+\mathbf{q})},n_{-\mathbf{k}};N_\mathbf{q}\rangle$
and
$|n_{+(\mathbf{k}+\mathbf{q})}+1,n_{-\mathbf{k}}-1;N_\mathbf{q}-1 \rangle$, respectively.
Thus, the only non-vanishing terms following from equation (\ref{H1}) read
\begin{widetext}
\begin{equation}
 \langle N_\mathbf{q}+1; n_{+\mathbf{k}}-1,n_{-(\mathbf{k}-\mathbf{q})}+1 |
 c_{-(\mathbf{k}-\mathbf{q})}^\dagger c_{+\mathbf{k}} a_\mathbf{q}^\dagger
 |n_{-(\mathbf{k}-\mathbf{q})},n_{+\mathbf{k}};N_\mathbf{q}\rangle
 = \sqrt{N_\mathbf{q}+1}\sqrt{n_{+\mathbf{k}}}\sqrt{1-n_{-(\mathbf{k}-\mathbf{q})}},
\end{equation}
\begin{equation}
 \langle N_\mathbf{q}-1; n_{-\mathbf{k}}-1, n_{+(\mathbf{k}+\mathbf{q})}+1 |
 c_{+(\mathbf{k}+\mathbf{q})}^\dagger c_{-\mathbf{k}} a_\mathbf{q}
 |n_{+(\mathbf{k}+\mathbf{q})},n_{-\mathbf{k}};N_\mathbf{q}\rangle
 = \sqrt{N_\mathbf{q}}\sqrt{n_{-\mathbf{k}}}\sqrt{1-n_{+(\mathbf{k}+\mathbf{q})}},
\end{equation}
\end{widetext}
and the respective rates are given by
\begin{eqnarray}
\nonumber && W^\mathrm{abs}_{\mathbf{k}\to \mathbf{k}+\mathbf{q}} =  \frac{\pi^2 e^2 v^2e^{-2\kappa_d z_0}}{\omega L_xL_y L_z}
\sin^2\left(\theta-\frac{\phi+\phi'}{2}\right)\\
\nonumber  && \times N_\mathbf{q} \left[1-n_{+(\mathbf{k}+\mathbf{q})}\right] n_{-\mathbf{k}} 
 \delta\left(\varepsilon_{\mathbf{k}+\mathbf{q}} + \varepsilon_{\mathbf{k}} -\hbar\omega\right),\\
 \label{wabs}
 \end{eqnarray}
 \begin{eqnarray}
 \nonumber && W^\mathrm{emi}_{\mathbf{k}\to \mathbf{k}-\mathbf{q}} = \frac{\pi^2 e^2 v^2e^{-2\kappa_d z_0}}{\omega L_xL_y L_z}
 \sin^2\left(\theta-\frac{\phi+\phi'}{2}\right)\\
\nonumber  &&\times \left(N_\mathbf{q} +1\right) n_{+\mathbf{k}} \left[1-n_{-(\mathbf{k}-\mathbf{q})}\right] 
 \delta\left(\hbar\omega -\varepsilon_{\mathbf{k}-\mathbf{q}} - \varepsilon_{\mathbf{k}} \right). \\
 \label{wemi}
\end{eqnarray}
It is convenient to reformulate the rates in terms of  $\theta$ and $\phi$ and
employ the explicit relations for electron energies as
\begin{equation}
\varepsilon_{\mathbf{k}}=\hbar v k, \quad \varepsilon_{\mathbf{k}\pm\mathbf{q}} =\hbar v\sqrt{k^2+q^2 \pm 2kq\cos(\theta-\phi)}.
\end{equation}
Equations (\ref{wabs}--\ref{wemi}) are then given by
\begin{eqnarray}
\nonumber && W^\mathrm{abs}_{\mathbf{k}\to \mathbf{k}+\mathbf{q}} =  \frac{\pi^2 e^2 v^2e^{-2\kappa_d z_0}}{\omega L_xL_y L_z}\\
\nonumber && \times \frac{1}{2}\left[1- \frac{q\cos(\theta-\phi)+k\cos(2\theta-2\phi)}{\sqrt{k^2+q^2+2kq\cos(\theta-\phi)}} \right] \\
\nonumber  && \times N_\mathbf{q} n_{-\mathbf{k}} \left[1-n_{+(\mathbf{k}+\mathbf{q})}\right] \\
 \nonumber && \times\delta\left(\hbar\omega -\hbar v\sqrt{k^2+q^2+2kq\cos(\theta-\phi)}-\hbar v k \right),\\
 \label{wabs2}
 \end{eqnarray}
 \begin{eqnarray}
 \nonumber && W^\mathrm{emi}_{\mathbf{k}\to \mathbf{k}-\mathbf{q}} = \frac{\pi^2 e^2 v^2e^{-2\kappa_d z_0}}{\omega L_xL_y L_z}\\
 \nonumber && \times \frac{1}{2}\left[1+ \frac{q\cos(\theta-\phi)-k\cos(2\theta-2\phi)}{\sqrt{k^2+q^2-2kq\cos(\theta-\phi)}} \right] \\
\nonumber  &&\times \left(N_\mathbf{q} +1\right) n_{+\mathbf{k}} \left[1-n_{-(\mathbf{k}-\mathbf{q})}\right] \\ 
 \nonumber && \times\delta\left(\hbar\omega -\hbar v\sqrt{k^2+q^2-2kq\cos(\theta-\phi)}-\hbar v k \right).\\
 \label{wemi2}
\end{eqnarray}

\section{S4. SPP attenuation}

Before calculating the SPP generation rate it is instructive to consider SPP attenuation assuming that the electrons are in a quasi-equilibrium state
characterized by the electron temperature $T_e$ and quasi-Fermi energy levels for electrons and holes, $E_{Fn}$ and $E_{Fp}$. 
(The both are positive numbers determined by optically excited electron and hole concentrations in graphene.)
The bias field is switched off at the moment, hence, we can set the x-axis along the SPP wave vector $\mathbf{q}$ without generality loss.
This allows for derivation of an explicit expression for the SPP decay rate.
In the Cartesian coordinates, the absorption and emission rates then read
\begin{eqnarray}
\nonumber && W^\mathrm{abs}_{\mathbf{k}\to \mathbf{k}+\mathbf{q}} = \frac{\pi^2 e^2 v^2}{\omega L_xL_y L_z}e^{-2\kappa_d z_0}\\
\nonumber && \times \frac{1}{2}\left[1- \frac{k_x(k_x+q)-k_y^2}{\sqrt{(k_x+q)^2+k_y^2}\sqrt{k_x^2+k_y^2}} \right] \\
\nonumber  &&\times N_\mathbf{q} \left[1-n_{+(\mathbf{k}+\mathbf{q})}\right] n_{-\mathbf{k}} 
 \delta\left(\varepsilon_{\mathbf{k}+\mathbf{q}} + \varepsilon_{\mathbf{k}} -\hbar\omega\right), \\
 \label{wabs0}
 \end{eqnarray}
 \begin{eqnarray}
 \nonumber && W^\mathrm{emi}_{\mathbf{k}\to \mathbf{k}-\mathbf{q}} = \frac{\pi^2 e^2 v^2}{\omega L_xL_y L_z}e^{-2\kappa_d z_0}\\
\nonumber  && \times \frac{1}{2}\left[1- \frac{k_x(k_x-q)-k_y^2}{\sqrt{(k_x-q)^2+k_y^2}\sqrt{k_x^2+k_y^2}} \right] \\
\nonumber  &&\times \left(N_\mathbf{q} +1\right) n_{+\mathbf{k}} \left[1-n_{-(\mathbf{k}-\mathbf{q})}\right] 
 \delta\left(\hbar\omega -\varepsilon_{\mathbf{k}-\mathbf{q}} - \varepsilon_{\mathbf{k}} \right). \\
 \label{wemi0}
\end{eqnarray}
The electron distribution functions can be written as
\begin{eqnarray}
 n_{+(\mathbf{k}+\mathbf{q})} & = & \frac{1}{1+\exp\left(\frac{\hbar v \sqrt{(k_x+q)^2+k_y^2} - E_{Fn}}{k_B T_e}\right)},\\
  n_{-\mathbf{k}} & = & \frac{1}{1+\exp\left(\frac{-\hbar v \sqrt{k_x^2+k_y^2} + E_{Fp}}{k_B T_e}\right)},\\
  n_{+\mathbf{k}} & = & \frac{1}{1+\exp\left(\frac{\hbar v \sqrt{k_x^2+k_y^2} - E_{Fn}}{k_B T_e}\right)},\\  
  n_{-(\mathbf{k}-\mathbf{q})} & = & \frac{1}{1+\exp\left(\frac{-\hbar v \sqrt{(k_x-q)^2+k_y^2} + E_{Fp}}{k_B T_e}\right)},
\end{eqnarray}
and $N_\mathbf{q}$ is to be found from the rate equation given by
\begin{equation}
 \frac{d N_\mathbf{q}}{d t}= \sum\limits_{\mathbf{k}} \left(W^\mathrm{emi}_{\mathbf{k}\to \mathbf{k}-\mathbf{q}}
 -  W^\mathrm{abs}_{\mathbf{k}\to \mathbf{k}+\mathbf{q}}\right).
\end{equation}
Summation can be changed to integration as
\begin{equation}
 \sum\limits_{\mathbf{k}} \to \int \frac{d k_x L_x}{2\pi}  \int \frac{d k_y L_y}{2\pi}, 
\end{equation}
where the limits of the integral over $k_y$ are from $-\infty$ to $+\infty$, and
the limits of the integral over $k_x$ are determined by the energy conservation
encoded in the $\delta$-functions of equations (\ref{wabs}) and (\ref{wemi}).
The $\delta$-functions can be transformed as
\begin{eqnarray}
 \nonumber && \delta\left(\hbar\omega -\varepsilon_{\mathbf{k}\pm\mathbf{q}} - \varepsilon_{\mathbf{k}} \right)=
 \frac{\delta\left(k_y-k_y^\pm\right)+\delta\left(k_y+k_y^\pm\right)}{2\hbar v\omega^2}\\
 && \times \frac{\lvert \omega^4-q^2 v^4 \left(2k_x\pm q \right)^2\rvert}{\sqrt{\omega^2-q^2v^2}\sqrt{\omega^2-\left(2k_x\pm q\right)^2v^2}},
 \label{deltaE}
\end{eqnarray}
 where
 \begin{equation}
  k_y^\pm= \frac{\sqrt{\omega^2-q^2v^2}}{2v\omega}\sqrt{\omega^2-(4v^2k_x^2\pm 4v^2qk_x +q^2 v^2)}.  
  \label{ky1}
 \end{equation}

 Equation (\ref{ky1}) can be reformulated as
 \begin{equation}
  k_y^\pm= \sqrt{\frac{\left[\omega^2 -v^2 q(q\pm 2k_x) \right]^2}{4v^2\omega^2}-k_x^2},
  \label{ky2}
 \end{equation}
 that makes it easier to write the following relations:
\begin{equation}
 \sqrt{k_x^2+(k_y^\pm)^2}=\frac{\lvert\omega^2 - v^2 q(q\pm 2k_x)\rvert}{2v\omega},
 \label{kxy}
\end{equation}
\begin{equation}
 \sqrt{(k_x\pm q)^2+(k_y^\pm)^2}=\frac{\lvert\omega^2 + v^2 q(q\pm 2k_x)\rvert }{2v\omega}.
 \label{kxyq}
\end{equation}
As the sum of equations (\ref{kxy}) and (\ref{kxyq}) must result in $\omega/v$ 
(the energy conservation is violated otherwise), hence,
the conditions 
\begin{eqnarray}
\frac{\left[\omega^2 -v^2 q(q+ 2k_x) \right]^2}{4v^2\omega^2}-k_x^2 \geq 0, \\
\frac{\left[\omega^2 -v^2 q(q- 2k_x) \right]^2}{4v^2\omega^2}-k_x^2 \geq 0,
\end{eqnarray}
determine the limits of the integral over $k_x$ for absorption and emission terms respectively.

Once integrated over $k_y$ the chirality factor takes the form
\begin{eqnarray}
 &&  \frac{1}{2}\left[1- \frac{k_x(k_x\pm q)-(k_y^\pm)^2}{\sqrt{(k_x\pm q)^2+k_y^2}\sqrt{k_x^2+k_y^2}} \right]=\\
 \nonumber && \frac{\lvert \omega^4-q^2 v^4 \left(q\pm 2k_x \right)^2\rvert +\omega^4 
 + (q\pm 2k_x)^2 \left(v^4q^2 - 2v^2\omega^2\right) }{2\lvert \omega^4-q^2 v^4 \left(q\pm 2k_x \right)^2\rvert}
\end{eqnarray}
with the denominator canceling out after multiplying by equation (\ref{deltaE}).

After some algebra the emission term can be written as
\begin{eqnarray}
 && \sum\limits_{\mathbf{k}} W^\mathrm{emi}_{\mathbf{k}\to \mathbf{k}-\mathbf{q}}=\int\limits_{\frac{q}{2}-\frac{\omega}{2v}}^{\frac{q}{2}+\frac{\omega}{2v}} d k_x \frac{e^2 v}{8\hbar\omega L_z}e^{-2\kappa_d z_0}\\
 \nonumber &&
 \frac{\lvert \omega^4-q^2 v^4 \left(q - 2k_x \right)^2\rvert +\omega^4 
 + (q - 2k_x)^2 \left(v^4q^2 - 2v^2\omega^2\right) }{\omega^2\sqrt{\omega^2-q^2v^2}\sqrt{\omega^2-\left(2k_x- q\right)^2v^2}}\\
 \nonumber &&
 \times \frac{N_\mathbf{q}+1}{1+\exp\left(\frac{\hbar \lvert\omega^2 - v^2 q(q- 2k_x)\rvert }{2 \omega k_B T_e}-\frac{E_{Fn}}{k_B T_e}\right)}  \\
 \nonumber && \times
 \left[1-\frac{1}{1+\exp\left(-\frac{\hbar \lvert\omega^2 + v^2 q(q- 2k_x)\rvert }{2\omega k_B T_e}+\frac{E_{Fp}}{k_B T_e}\right)} \right],
\end{eqnarray}
whereas the absorption term reads
\begin{eqnarray}
 && \sum\limits_{\mathbf{k}} W^\mathrm{abs}_{\mathbf{k}\to \mathbf{k}+\mathbf{q}}=\int\limits_{-\frac{q}{2}-\frac{\omega}{2v}}^{-\frac{q}{2}+\frac{\omega}{2v}} d k_x \frac{e^2 v}{8\hbar\omega L_z}e^{-2\kappa_d z_0}\\
 \nonumber &&
 \frac{\lvert \omega^4-q^2 v^4 \left(q + 2k_x \right)^2\rvert +\omega^4 
 + (q + 2k_x)^2 \left(v^4q^2 - 2v^2\omega^2\right) }{\omega^2 \sqrt{\omega^2-q^2v^2}\sqrt{\omega^2-\left(2k_x+q\right)^2v^2}}\\
 \nonumber &&
 \times \frac{N_\mathbf{q}}{1+\exp\left(-\frac{\hbar \lvert\omega^2 - v^2 q(q+ 2k_x)\rvert }{2 \omega k_B T_e}+\frac{E_{Fp}}{k_B T_e}\right)}  \\
 \nonumber && \times
 \left[1-\frac{1}{1+\exp\left(\frac{\hbar \lvert\omega^2 + v^2 q(q+2k_x)\rvert }{2\omega k_B T_e}-\frac{E_{Fn}}{k_B T_e}\right)} \right].
\end{eqnarray}
Substituting $k_x\to k_x+q$ in the emission term we set the limits of integration the same as in the absorption term.
Besides, we transform the Fermi-Dirac distributions as
\begin{eqnarray}
 && \frac{1}{1+\exp\left(\frac{\hbar \lvert\omega^2 + v^2 q(q+ 2k_x)\rvert }{2 \omega k_B T_e}-\frac{E_{Fn}}{k_B T_e}\right)}=\\
 \nonumber && \exp\left(\frac{E_{Fn}}{k_B T_e}-\frac{\hbar \lvert\omega^2 + v^2 q(q+2k_x)\rvert }{2\omega k_B T_e}\right)\\
 \nonumber && \times 
 \left[1 -\frac{1}{1+\exp\left(\frac{\hbar \lvert\omega^2 + v^2 q(q+2k_x)\rvert }{2\omega k_B T_e}-\frac{E_{Fn}}{k_B T_e}\right)} \right],
\end{eqnarray}
\begin{eqnarray}
 && 1 -\frac{1}{1+\exp\left(-\frac{\hbar \lvert\omega^2 - v^2 q(q+2k_x)\rvert }{2\omega k_BT_e}+\frac{E_{Fp}}{k_BT_e}\right)}=\\
 \nonumber && 
 \frac{\exp\left(-\frac{\hbar \lvert\omega^2 - v^2 q(q+2k_x)\rvert }{2\omega k_BT_e}+\frac{E_{Fp}}{k_BT_e}\right) }{1+\exp\left(-\frac{\hbar \lvert\omega^2 - v^2 q(q+2k_x)\rvert }{2\omega k_BT_e}+\frac{E_{Fp}}{k_BT_e}\right)}.
\end{eqnarray}
The sum of the emission and absorption terms then reads
\begin{eqnarray}
 \nonumber && \frac{d N_\mathbf{q}}{d t} =\int\limits_{-\frac{q}{2}-\frac{\omega}{2v}}^{-\frac{q}{2}+\frac{\omega}{2v}} d k_x \frac{e^2 v}{8\hbar\omega L_z}e^{-2\kappa_d z_0}\\
 \nonumber &&
 \frac{\lvert \omega^4-q^2 v^4 \left(q + 2k_x \right)^2\rvert +\omega^4 
 + (q + 2k_x)^2 \left(v^4q^2 - 2v^2\omega^2\right) }{\omega^2 \sqrt{\omega^2-q^2v^2}\sqrt{\omega^2-\left(2k_x+q\right)^2v^2}}\\
 \nonumber &&
  \times \frac{\left(N_\mathbf{q}+1\right)e^{\frac{\Delta E_F-\hbar\omega}{k_B T_e}}- N_\mathbf{q}}{1+\exp\left(-\frac{\hbar \lvert\omega^2 - v^2 q(q+ 2k_x)\rvert }{2 \omega k_B T_e}+\frac{E_{Fp}}{k_B T_e}\right)}  \\
   && \times
 \left[1-\frac{1}{1+\exp\left(\frac{\hbar \lvert\omega^2 + v^2 q(q+2k_x)\rvert }{2\omega k_B T_e}-\frac{E_{Fn}}{k_B T_e}\right)} \right],
 \label{rate-eq}
\end{eqnarray}
where $\Delta E_F = E_{Fn} + E_{Fp}$.
We look for the solution of equation (\ref{rate-eq}) in the form $N_\mathbf{q}=N_\mathbf{q}^{(0)}+N_\mathbf{q}^{(1)}$,
where
\begin{equation}
 N_\mathbf{q}^{(0)} = \frac{1}{\mathrm{e}^{\frac{\hbar\omega-\Delta E_F}{k_B T_0}}-1}
 \label{Nq0}
\end{equation}
is the Bose-Einstein distribution for SPP with $\Delta E_F$ sometimes referred to as the chemical
potential of radiation.
If we have no bias heating up electrons, then $T_e=T_0$, and
making use the following relation
\begin{equation}
 \left(N_\mathbf{q}^{(0)}+1\right)\mathrm{e}^{\frac{\Delta E_F-\hbar\omega}{k_B T_e}}- N_\mathbf{q}^{(0)}=0,
 \label{eq-relation}
\end{equation}
the rate equation then takes the form
\begin{equation}
 \frac{d N_\mathbf{q}^{(1)}}{d t}=-\frac{N_\mathbf{q}^{(1)}}{\tau_\mathbf{q}},
\end{equation}
where
\begin{eqnarray}
 \nonumber && \frac{1}{\tau_\mathbf{q}}= \int\limits_{-\frac{q}{2}-\frac{\omega}{2v}}^{-\frac{q}{2}+\frac{\omega}{2v}} d k_x \frac{e^2 v}{8\hbar\omega L_z}e^{-2\kappa_d z_0}\\
 \nonumber &&
 \frac{\lvert \omega^4-q^2 v^4 \left(2k_x+q \right)^2\rvert +\omega^4 
 + (2k_x+q)^2 \left(v^4q^2 - 2v^2\omega^2\right) }{\omega^2  \sqrt{\omega^2-q^2v^2}\sqrt{\omega^2-\left(2k_x+q\right)^2v^2}}\\
 \nonumber &&
  \times \frac{1- e^{\frac{\Delta E_F-\hbar\omega}{k_B T_0}}}{1+\exp\left(-\frac{\hbar \lvert\omega^2 - v^2 q(2k_x+q)\rvert }{2 \omega k_B T_0}+\frac{E_{Fp}}{k_B T_0}\right)}  \\
   && \times
 \left[1-\frac{1}{1+\exp\left(\frac{\hbar \lvert\omega^2 + v^2 q(2k_x+q)\rvert }{2\omega k_B T_0}-\frac{E_{Fn}}{k_B T_0}\right)} \right].
 \label{tau}
\end{eqnarray}
Here, $\tau_\mathbf{q}$ is the SPP relaxation time due to the absorption in graphene.
Making use of the limits for $k_x$ the integrand can be simplified, and the result reads
\begin{eqnarray}
 \nonumber && \frac{1}{\tau_\mathbf{q}}= \frac{e^2 ve^{-2\kappa_d z_0}}{4\hbar\omega L_z}\int\limits_{-\frac{q}{2}-\frac{\omega}{2v}}^{-\frac{q}{2}+\frac{\omega}{2v}} d k_x \sqrt{\frac{\omega^2-\left(2k_x+q\right)^2v^2}{\omega^2-q^2v^2}}\\
 \nonumber &&
  \times \frac{1- e^{\frac{\Delta E_F-\hbar\omega}{k_B T_0}}}{1+\exp\left(-\frac{\hbar \lvert\omega^2 - v^2 q(2k_x+q)\rvert }{2 \omega k_B T_0}+\frac{E_{Fp}}{k_B T_0}\right)}  \\
   && \times
 \left[1-\frac{1}{1+\exp\left(\frac{\hbar \lvert\omega^2 + v^2 q(2k_x+q)\rvert }{2\omega k_B T_0}-\frac{E_{Fn}}{k_B T_0}\right)} \right].
 \label{tau-simpler}
\end{eqnarray}

At $T_0=0$ and $E_{Fn,p}=0$ equation (\ref{tau-simpler}) takes the form of {\bf Eq. (A9)} in the main text.




\begin{figure*}
 \includegraphics[width=0.7\textwidth]{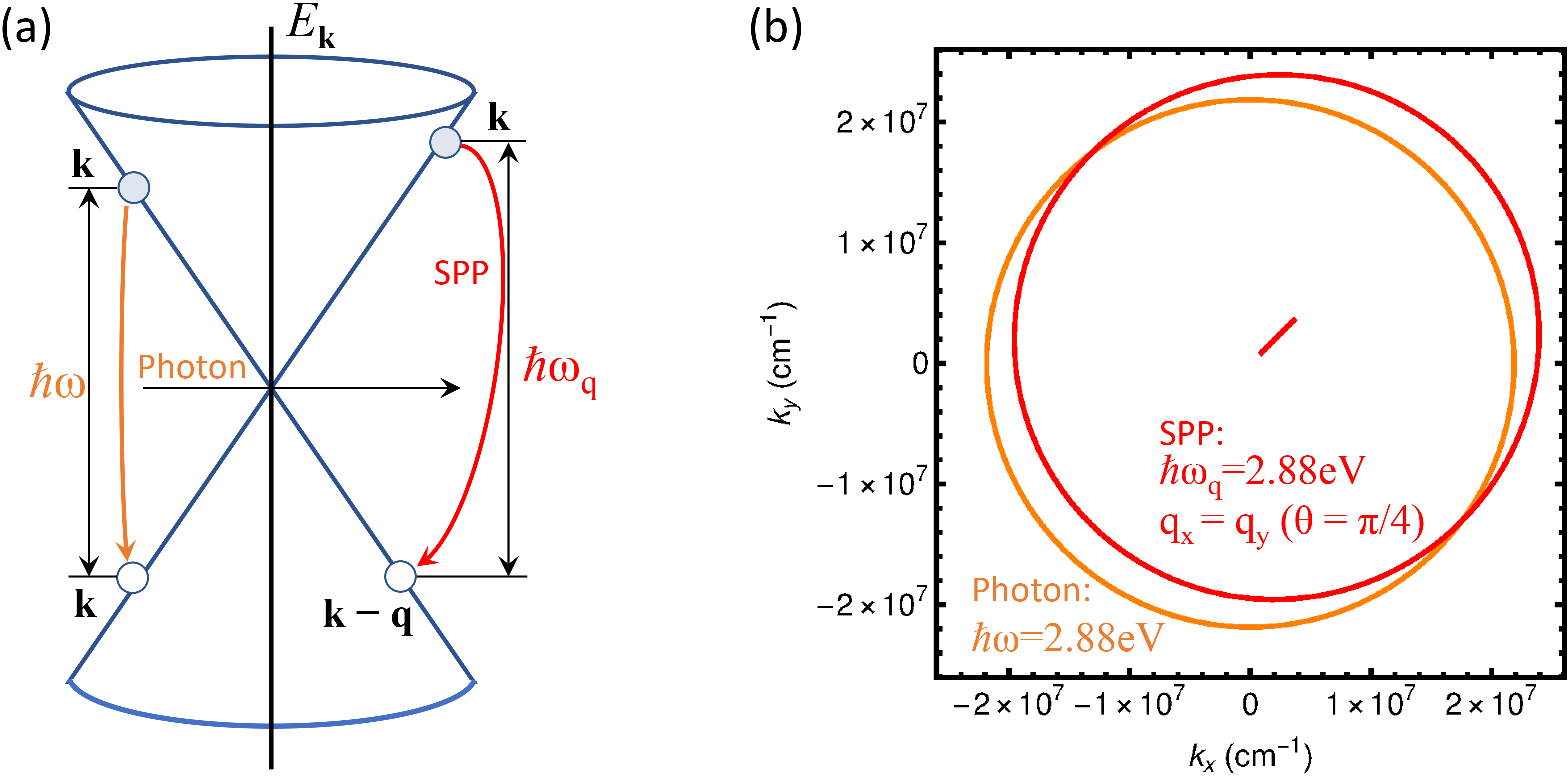}
 \caption{
Panel (a) illustrates the interband transitions responsible for photon and SPP emission: The photon transitions are direct, while
there is a momentum transfer in the case of SPPs.
Panel (b) displays the pairs of the electron wave vectors $k_x$ and $k_y$ satisfying the energy and momentum conservation 
in the SPP (red) and photon (orange) emission/absorption processes for a given excitation energy and selected direction of $\mathbf{q}$.
In the special case, where $\theta=\phi$ ($\mathbf{q}$ and $\mathbf{k}$ are collinear, see Eq. (\ref{delta-pm})),
the conservation condition reads $v|k-q|+v k=\omega$ so that it has two solutions 
(i) $2vk-vq=\omega$ for $k>q$, and (ii) $vq=\omega$ for $k<q$ independent of $k$.
The latter is represented by the short line in the middle of the plot in panel and responsible for the nesting effect in the SPP absorption, see {\bf Fig. 2}
in the main text.
This collinear solution does not exist for photons, where $k_x$ and $k_y$ satisfy 
the simple equation $2\hbar v k=\hbar\omega$.
}
\end{figure*}

\section{S5. Biased electrons}

To excite SPPs we need energy and momentum. Both can be acquired from an external electric field.
We consider an electric field applied along x-axis (${\cal E}_x$), whereas the direction of $\mathbf{q}$ is arbitrary.
The linear-response Boltzmann equation for electrons
within the relaxation rate approximation can be written as
\begin{equation}
  \pm e{\cal E}_x v_x\frac{d f_{\pm \mathbf{k}}^{(0)}}{d E_{\pm k}} = -\frac{f_{\pm \mathbf{k}}^{(1)}}{\tau_k},
 \label{B}
\end{equation}
where $\tau_k$ is the momentum relaxation time, 
$v_x=v\cos\phi$, $v_y=v\sin\phi$ are the $x$ and $y$ components of the electron velocity,
$f_{\pm \mathbf{k}}^{(1)}$ is the ${\cal E}_x$--dependent non-equilibrium addition to the Fermi-Dirac distribution function $f_{\pm \mathbf{k}}^{(0)}$ written explicitly as
\begin{eqnarray}
  f_{\pm\mathbf{k}}^{(0)} & = & \frac{1}{1+\exp\left(\frac{\pm \hbar v k \mp E_{Fn(p)}}{k_B T_e}\right)},\\  
 f_{\pm\mathbf{k}}^{(1)} & =  & \pm e {\cal E}_x v_x \tau_k
 \left(-\frac{d f_{\pm \mathbf{k}}^{(0)}}{dE_{\pm k}}\right)\\
 \nonumber & = &  \pm \frac{e{\cal E}_x v_x \tau_k}{4k_B T_e\cosh^2\left(\frac{\pm \hbar v k \mp E_{Fn(p)}}{2k_B T_e}\right)}.
\end{eqnarray}

It is useful to find the relation between electron mobility and microscopic disorder.
The momentum relaxation time depends on the disorder potential and can be calculated by means of the Fermi's golden-rule.
The assumption of the long-range disorder makes it possible to introduce
the electron mobility independent of the carrier concentration. Indeed, setting $\tau_k=\hbar v k \gamma $, 
$E_{Fn}=E_F$, $E_{Fp}=0$, $k_BT_e \ll E_F$, the current density reads
\begin{eqnarray}
 j_x & = &  g_{sv}\int\frac{d^2 k}{(2\pi)^2}  e^2{\cal E}_x v_x^2 \tau_k 
 \left(-\frac{d f^{(0)}_{+\mathbf{k}}}{dE_{+k}}\right)\\
 & = & \frac{e^2 \gamma E_F^2}{\pi \hbar^2}{\cal E}_x \\
 & = & e^2 \gamma v^2 n {\cal E}_x\\
 & = & e\mu n {\cal E}_x,
\end{eqnarray}
where $n=E_F^2/\pi\hbar^2 v^2$ is the equilibrium electron concentration at zero temperature,
$g_{sv}=4$ is the spin/valley degeneracy, and $\mu=e\gamma v^2$ is the electron mobility.
The latter is what we need to relate the microscopic and macroscopic parameters.
Using the relation $v_{x,y}=vk_{x,y}/k$ the non-equilibrium addition to the electron distribution can be then written as
\begin{eqnarray}
 f_{\pm\mathbf{k}}^{(1)} & = &  \pm \frac{\mu{\cal E}_x}{v}\frac{\hbar v k \cos\phi}{4k_BT_e\cosh^2\left(\frac{\pm \hbar v k \mp E_{Fn(p)}}{2k_BT_e}\right)},
\end{eqnarray}
where $\mu {\cal E}_x$ has the meaning of the drift velocity.
It is also instructive to write $f_{\pm(\mathbf{k}\pm\mathbf{q})}^{(0)}$ and $f_{\pm(\mathbf{k}\pm\mathbf{q})}^{(1)}$ explicitly as
\begin{eqnarray}
 && f_{\pm(\mathbf{k}\pm\mathbf{q})}^{(0)} =  \\
 \nonumber &&\frac{1}{1+\exp\left(\frac{\pm \hbar v \sqrt{k^2+q^2 \pm 2kq\cos(\theta-\phi)} \mp E_{Fn(p)}}{k_BT_e}\right)},
\end{eqnarray}
\begin{eqnarray}
 && f_{\pm(\mathbf{k}\pm\mathbf{q})}^{(1)} =  \\
 \nonumber && \pm \frac{\mu{\cal E}_x}{v}\frac{\hbar v (k \cos\phi \pm q\cos\theta)}{4k_BT_e\cosh^2\left(\frac{\pm \hbar v \sqrt{k^2+q^2 \pm 2kq\cos(\theta-\phi)} \mp E_{Fn(p)}}{2k_BT_e}\right)}.
 \end{eqnarray}
Hence, the electron occupations in equations (\ref{wabs2}--\ref{wemi2}) can be written explicitly as
\begin{eqnarray}
 n_{+(\mathbf{k}+\mathbf{q})} & = & f_{+(\mathbf{k}+\mathbf{q})}^{(0)}+f_{+(\mathbf{k}+\mathbf{q})}^{(1)},\\
 n_{-\mathbf{k}} & = & f_{-\mathbf{k}}^{(0)}+f_{-\mathbf{k}}^{(1)},\\  
  n_{+\mathbf{k}} & = & f_{+\mathbf{k}}^{(0)}+f_{+\mathbf{k}}^{(1)},\\  
  n_{-(\mathbf{k}-\mathbf{q})} & = & f_{-(\mathbf{k}-\mathbf{q})}^{(0)}+f_{-(\mathbf{k}-\mathbf{q})}^{(1)}.
\end{eqnarray}

\section{S6. SPP generation}

The total SPP occupation reads $N_\mathbf{q}= N_\mathbf{q}^{(0)} + N_\mathbf{q}^{(1)}$,
where $N_\mathbf{q}^{(0)}$ is given by equation (\ref{Nq0}), and $N_\mathbf{q}^{(1)}$ is
the non-equilibrium term to be determined.

The $\delta$-functions in equations (\ref{wabs2}--\ref{wemi2}) can be transformed as 
\begin{eqnarray}
 \nonumber && \delta\left(\hbar\omega -\hbar v\sqrt{k^2+q^2\pm 2kq\cos(\theta-\phi)}-\hbar v k \right)= \\
 \nonumber && \frac{1}{\hbar v}\frac{\delta(k-k_\pm)\sqrt{k^2+q^2\pm 2kq\cos(\theta-\phi)}}{\sqrt{k^2+q^2\pm 2kq\cos(\theta-\phi)}+k\pm q\cos(\theta-\phi)},\\
 \label{delta-pm}
\end{eqnarray}
where
\begin{equation}
 \label{k-pm}
 k_\pm =\frac{1}{2v} \frac{\omega^2 - q^2 v^2 }{\omega\pm q v \cos(\theta-\phi)}\geq 0.
\end{equation}

Now, we are ready to find $N_\mathbf{q}^{(1)}$ in a steady state.
The difference between SPP emission and absorption rates reads
\begin{eqnarray}
 \nonumber && \sum\limits_{\mathbf{k}} \left(W^\mathrm{emi}_{\mathbf{k}\to \mathbf{k}-\mathbf{q}}
 -  W^\mathrm{abs}_{\mathbf{k}\to \mathbf{k}+\mathbf{q}}\right) =\\
 \nonumber && \int \frac{d k_x L_x}{2\pi}  \int \frac{d k_y L_y}{2\pi}
  \left(W^\mathrm{emi}_{\mathbf{k}\to \mathbf{k}-\mathbf{q}}
 -  W^\mathrm{abs}_{\mathbf{k}\to \mathbf{k}+\mathbf{q}}\right) = \\
 \nonumber && \frac{L_x L_y}{(2\pi)^2} \int\limits_0^{2\pi}d\phi \int\limits_0^\infty dk k  \left(W^\mathrm{emi}_{\mathbf{k}\to \mathbf{k}-\mathbf{q}}
 -  W^\mathrm{abs}_{\mathbf{k}\to \mathbf{k}+\mathbf{q}}\right) \\
 &&= G_\mathbf{q} - \frac{N_\mathbf{q}^{(1)}}{\tau_\mathbf{q}}.
\end{eqnarray}
The integral over $k$ is trivial due to the $\delta$-functions (\ref{delta-pm}). 
After lengthy transformations the integral over $\phi$ takes the form
\begin{eqnarray}
 \nonumber && \frac{e^2  e^{-2\kappa_d z_0}}{16 \hbar\omega L_z}\int\limits_0^{2\pi}d\phi 
\left\{ \frac{\omega^2 (\omega^2-q^2 v^2)\sin^2(\theta-\phi)}{\left[\omega - q v \cos(\theta-\phi)\right]^3} \right. \\
 \nonumber && \times \left(N_\mathbf{q} +1\right) \left[ n_{+\mathbf{k}} \left( 1-n_{-(\mathbf{k}-\mathbf{q})}\right)\right]_{k=k_-}  \\ 
\nonumber  && \left. 
 - \frac{\omega^2 (\omega^2-q^2 v^2)\sin^2(\theta-\phi)}{\left[\omega + q v \cos(\theta-\phi)\right]^3}
 N_\mathbf{q} \left[n_{-\mathbf{k}} \left(1-n_{+(\mathbf{k}+\mathbf{q})}\right)\right]_{k=k_+}\right\}\\
 && = G_\mathbf{q} - \frac{N_\mathbf{q}^{(1)}}{\tau_\mathbf{q}},
\end{eqnarray}
where $\tau_\mathbf{q}$ is the SPP relaxation time and $G_\mathbf{q}$ is the SPP generation rate.


The SPP decay rate is given by
\begin{eqnarray}
 \nonumber && -\frac{1}{\tau_\mathbf{q}} = \frac{e^2  e^{-2\kappa_d z_0}}{16 \hbar\omega L_z}\int\limits_0^{2\pi}d\phi 
\left\{ \frac{\omega^2 (\omega^2-q^2 v^2)\sin^2(\theta-\phi)}{\left[\omega - q v \cos(\theta-\phi)\right]^3} \right.\\
 \nonumber && 
 \times \left[f_{+\mathbf{k}}^{(0)}\left(1-f_{-(\mathbf{k}-\mathbf{q})}^{(0)}\right)+f_{+\mathbf{k}}^{(1)}\left(1-f_{-(\mathbf{k}-\mathbf{q})}^{(0)}\right)
 \right. \\
 \nonumber && \left. 
 - f_{+\mathbf{k}}^{(0)}f_{-(\mathbf{k}-\mathbf{q})}^{(1)}- f_{+\mathbf{k}}^{(1)}f_{-(\mathbf{k}-\mathbf{q})}^{(1)} \right]_{k=k_-} \\
 \nonumber  &&  
 -\frac{\omega^2 (\omega^2-q^2 v^2)\sin^2(\theta-\phi)}{\left[\omega + q v \cos(\theta-\phi)\right]^3} \\
 \nonumber && \times \left[
 \left(1-f_{+(\mathbf{k}+\mathbf{q})}^{(0)}\right)f_{-\mathbf{k}}^{(0)}+
 \left(1-f_{+(\mathbf{k}+\mathbf{q})}^{(0)}\right)f_{-\mathbf{k}}^{(1)} \right. \\
  &&\left.   \left. -f_{+(\mathbf{k}+\mathbf{q})}^{(1)}f_{-\mathbf{k}}^{(0)}
- f_{+(\mathbf{k}+\mathbf{q})}^{(1)}f_{-\mathbf{k}}^{(1)} \right]_{k=k_+} \right\}
\label{tau-total}
\end{eqnarray}
Equation (\ref{tau-total}) is used to plot the solid curve in {\bf Fig. 2(a)} of the main text along with
the approximate expression for $1/\tau_\mathbf{q}$.

\begin{figure}
 \includegraphics[width=0.8\columnwidth]{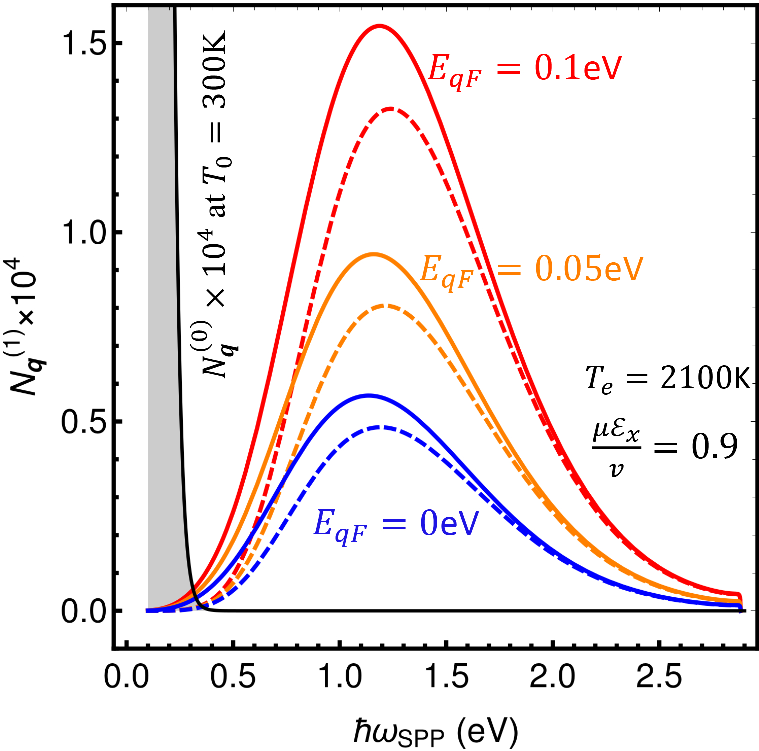}
 \caption{Non-equilibrium SPP distribution function compared to the thermal SPP bath distribution (gray shaded area) 
 at a finite quasi-Fermi energy ($E_{F(n,p)}=E_{qF}$) occured, e.g. when graphene electrons are in addition heated by THz radiation.
 Similar to {\bf Fig. 3} in the main text, solid curves represent the full solution, while dashed curves correspond to the approximate model.}
\end{figure}

The SPP generation rate is  given by
\begin{eqnarray}
 \nonumber && G_\mathbf{q}= \frac{e^2  e^{-2\kappa_d z_0}}{16 \hbar\omega L_z}\int\limits_0^{2\pi}d\phi 
\left\{ \frac{\omega^2 (\omega^2-q^2 v^2)\sin^2(\theta-\phi)}{\left[\omega - q v \cos(\theta-\phi)\right]^3} \right.\\
 \nonumber && 
 \times \left[f_{+\mathbf{k}}^{(0)}\left(1-f_{-(\mathbf{k}-\mathbf{q})}^{(0)}\right)+f_{+\mathbf{k}}^{(1)}\left(1-f_{-(\mathbf{k}-\mathbf{q})}^{(0)}\right)
 \right. \\
 \nonumber && \left. 
 - f_{+\mathbf{k}}^{(0)}f_{-(\mathbf{k}-\mathbf{q})}^{(1)}- f_{+\mathbf{k}}^{(1)}f_{-(\mathbf{k}-\mathbf{q})}^{(1)} \right]_{k=k_-} \\
 \nonumber  && \times \left(N_\mathbf{q}^{(0)} +1\right)
 -\frac{\omega^2 (\omega^2-q^2 v^2)\sin^2(\theta-\phi)}{\left[\omega + q v \cos(\theta-\phi)\right]^3} \\
 \nonumber &&  \times \left[
 \left(1-f_{+(\mathbf{k}+\mathbf{q})}^{(0)}\right)f_{-\mathbf{k}}^{(0)}+
 \left(1-f_{+(\mathbf{k}+\mathbf{q})}^{(0)}\right)f_{-\mathbf{k}}^{(1)} \right. \\
  &&\left.  \left. -f_{+(\mathbf{k}+\mathbf{q})}^{(1)}f_{-\mathbf{k}}^{(0)}
- f_{+(\mathbf{k}+\mathbf{q})}^{(1)}f_{-\mathbf{k}}^{(1)} \right]_{k=k_+}  N_\mathbf{q}^{(0)} \right\} \
\label{gen-total}
\end{eqnarray}
Equation (\ref{gen-total}) is used to plot the solid curves in {\bf Fig. 3} of the main text along with
the approximate expression for $G_\mathbf{q}$.

The terms linear in ${\cal E}_x$ vanish in Eq. (\ref{gen-total}), and the generation rate can be further simplified assuming 
$N_\mathbf{q}^{(0)}=0$ at low $T_0$ and $E_{F(n,p)}=0$. The result explicitly reads
\begin{eqnarray}
  \nonumber G_\mathbf{q} &= & \frac{e^2  e^{-2\kappa_d z_0}}{16 \hbar\omega L_z}\left(\frac{\mu{\cal E}_x}{v}\right)^2
  \left(\frac{\hbar v}{4T} \right)^2 \int\limits_0^{2\pi}d\phi \\
  \nonumber &&  \times \frac{\omega^2 (\omega^2-q^2 v^2)\sin^2(\theta-\phi)}{\left[\omega - q v \cos(\theta-\phi)\right]^3}
  \frac{ k_- \cos\phi}{\cosh^2\left(\frac{\hbar v k_- }{2k_BT_e}\right)}\\
  &&
  \times \frac{k_- \cos\phi - q\cos\theta}{\cosh^2\left(\frac{\hbar v \sqrt{k_-^2+q^2 - 2k_-q\cos(\theta-\phi)}}{2k_BT_e}\right)},
  \label{Gq0}
\end{eqnarray}
where $k_-$ is given by equation (\ref{k-pm}). 

Note that
\begin{eqnarray}
\nonumber && \sqrt{k_-^2+q^2 - 2k_-q\cos(\theta-\phi)} = \\
\nonumber && \frac{1}{2v} \frac{\omega^2 + q^2 v^2 - 2\omega qv\cos(\theta-\phi) }{\omega - q v \cos(\theta-\phi)},
\end{eqnarray}
that results in the energy conservation 
\begin{equation}
v\sqrt{k_-^2+q^2 - 2k_-q\cos(\theta-\phi)}+ vk_- = \omega. 
\label{aux1}
\end{equation}
In the limit $T_e=0$ the $\cosh^{-2}$--functions can be represented by two $\delta$--functions,
which do not overlap. The generation rate therefore vanishes at $T_e=0$ unless  $E_{F(n,p)}\neq 0$.
The generation rate is then maximized at $\hbar\omega=E_{Fp}+E_{Fn}$.

\section{S7. Photon emission}

In this section, we derive the total photon emission rate from graphene placed into vacuum.
Theory of far-field photon emission from 2D materials can be found in the book \cite{vasko1998electronic}
(see Chapter 6). We adopt the total emission rate equation, which in our notation can be written as
\begin{eqnarray}
\nonumber \frac{\Gamma_0}{L_x L_y} & = & \int\frac{d^3Q }{(2\pi)^3} \frac{e^2}{\omega}\int d^2 k \lvert\mathbf{e}_\mathbf{E}\cdot \mathbf{v}(\mathbf{k},\mathbf{q})\rvert^2\\
\nonumber && \times
 \delta\left(\hbar\omega -\varepsilon_{\mathbf{k}-\mathbf{q}} - \varepsilon_{\mathbf{k}} \right)
 n_{+\mathbf{k}} \left[1-n_{-(\mathbf{k}-\mathbf{q})}\right].\\
 \label{totrate}
\end{eqnarray}
Here, $\mathbf{e}_\mathbf{E}$ is the polarization vector, and $\omega=cQ$ is the photon dispersion with the wave vector $Q=\sqrt{q^2+q_z^2}$,
where $q$ and $q_z$ are the in-plane and out-of-plane components, respectively.

The interband transition matrix elements are calculated in a similar way as for SPP, however,
the polarization vector does not coincide with the wave vector and remains normal to propagation direction.
In fact, two polarizations (``p'' for ``parallel'' and ``s'' for ``senkrecht'') are possible,
and we have to sum up the respective contributions in equation (\ref{totrate}).
We introduce the azimuthal and polar polarization angles as $\phi_\mathbf{E}$ and $\beta_\mathbf{E}$, respectively,
that results in the matrix elements given by
\begin{eqnarray}
 && \lvert\mathbf{e}_\mathbf{E}\cdot \mathbf{v}(\mathbf{k},\mathbf{q})\rvert^2=\\
 \nonumber &&
 \frac{v^2 \sin^2\beta_\mathbf{E}}{2}\left[1+ \frac{q\cos(\theta+\phi-2\phi_\mathbf{E})-k\cos(2\phi-2\phi_\mathbf{E})}{\sqrt{k^2+q^2-2kq\cos(\theta-\phi)}} \right].
\end{eqnarray}
The integral over the electronic states then reads
\begin{eqnarray}
&& \nonumber \int d^2 k \lvert\mathbf{e}_\mathbf{E}\cdot \mathbf{v}(\mathbf{k},\mathbf{q})\rvert^2
 \delta\left(\hbar\omega -\varepsilon_{\mathbf{k}-\mathbf{q}} - \varepsilon_{\mathbf{k}} \right)\\
 \nonumber && \times n_{+\mathbf{k}} \left[1-n_{-(\mathbf{k}-\mathbf{q})}\right]  = \frac{\sin^2\beta_\mathbf{E}}{4\hbar}\\
 \nonumber && \times \int \limits_0^{2\pi} d\phi 
 \frac{(\omega^2-q^2 v^2)\left[\omega\sin(\phi-\phi_\mathbf{E})-qv\sin(\theta-\phi_\mathbf{E})\right]^2}{\left[\omega - q v \cos(\theta-\phi)\right]^3} \\
 \nonumber && \times n_{+\mathbf{k}} \left[1-n_{-(\mathbf{k}-\mathbf{q})}\right]_{k=k_-},
\end{eqnarray}
where $k_-$ is given by equation (\ref{k-pm}).
In contrast to SPP, we can assume $\omega \gg q v$ for photons (because $v\gg c$), and the integral takes the form
\begin{eqnarray}
&& \nonumber \int d^2 k \lvert\mathbf{e}_\mathbf{E}\cdot \mathbf{v}(\mathbf{k},\mathbf{q})\rvert^2
 \delta\left(\hbar\omega -\varepsilon_{\mathbf{k}-\mathbf{q}} - \varepsilon_{\mathbf{k}} \right)\\
 \nonumber && \times n_{+\mathbf{k}} \left[1-n_{-(\mathbf{k}-\mathbf{q})}\right] \\
 \nonumber && = \sin^2\beta_\mathbf{E} \frac{\omega}{4\hbar} \int \limits_0^{2\pi} d\phi 
 \sin^2(\phi-\phi_\mathbf{E})
 n_{+\mathbf{k}} \left[1-n_{-(\mathbf{k}-\mathbf{q})}\right]_{k=\frac{\omega}{2v}}.
\end{eqnarray}
In the case of p-polarized emission, the vectors $\mathbf{Q}$ and $\mathbf{E}$ are in the same plane and orthogonal, so that
$\phi_\mathbf{E}=\theta$ (here, $\theta$ is the the azimuthal angle of $\mathbf{Q}$, as in the previous sections), and
$\beta_\mathbf{E}=\beta+\pi/2$ (here, $\beta$ is the polar angle of $\mathbf{Q}$). The integral takes the form
\begin{eqnarray}
 \nonumber && \sin^2\beta_\mathbf{E} \frac{\omega}{4\hbar} \int \limits_0^{2\pi} d\phi 
 \sin^2(\phi-\phi_\mathbf{E})
 n_{+\mathbf{k}} \left[1-n_{-(\mathbf{k}-\mathbf{q})}\right]_{k=\frac{\omega}{2v}}\\
 \nonumber && = \cos^2\beta\frac{\omega}{4\hbar} \int \limits_0^{2\pi} d\phi 
 \sin^2(\phi-\theta)
 n_{+\mathbf{k}} \left[1-n_{-(\mathbf{k}-\mathbf{q})}\right]_{k=\frac{\omega}{2v}}.\\
 \label{p-pol}
\end{eqnarray}
In the case of s-polarized emission, the vector $\mathbf{E}$ is normal to  $\mathbf{Q}$ but parallel to the emitting surface, so that
$\phi_\mathbf{E}=\theta+\pi/2$, and $\beta_\mathbf{E}=\pi/2$. The integral takes the form
\begin{eqnarray}
 \nonumber && \sin^2\beta_\mathbf{E} \frac{\omega}{4\hbar} \int \limits_0^{2\pi} d\phi 
 \sin^2(\phi-\phi_\mathbf{E})
 n_{+\mathbf{k}} \left[1-n_{-(\mathbf{k}-\mathbf{q})}\right]_{k=\frac{\omega}{2v}}\\
 \nonumber && = \frac{\omega}{4\hbar} \int \limits_0^{2\pi} d\phi 
 \cos^2(\phi-\theta)
 n_{+\mathbf{k}} \left[1-n_{-(\mathbf{k}-\mathbf{q})}\right]_{k=\frac{\omega}{2v}}.\\
 \label{s-pol}
\end{eqnarray}

\begin{figure}
 \includegraphics[width=\columnwidth]{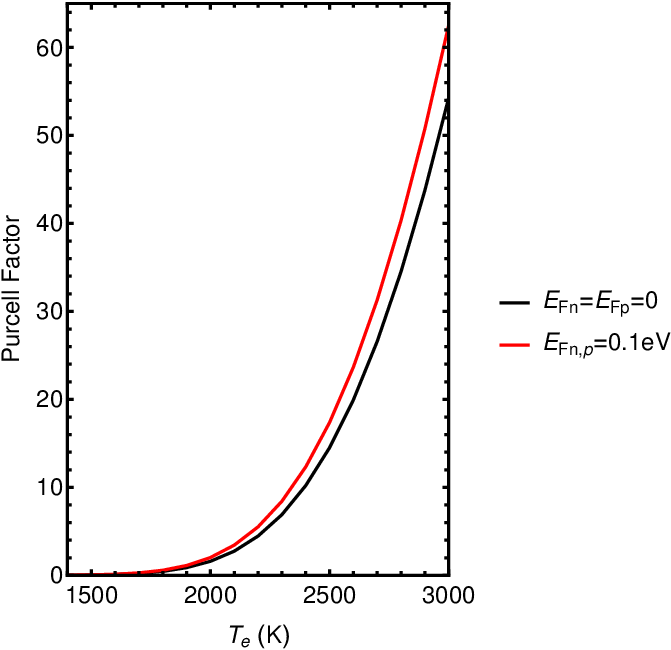}
 \caption{The Purcell factor, defined as the ratio of SPP to photon emission rates,
 does not show considerable enhancement at a finite quasi-Fermi energy $E_{F(n,p)}=0.1$eV,
 because increasing the number of electron-hole pairs proportionally enhances both photon and SPP emission rates. }
\end{figure}

In equation (\ref{totrate}) we sum up both emission channels (\ref{p-pol}) and (\ref{s-pol}), and then changing the integration
over $d^3 Q$ to the integration over $d\omega$ we arrive at the following explicit expression for the total photon emission
\begin{eqnarray}
\nonumber \frac{\Gamma_0}{L_x L_y} & = & \frac{4}{3}\frac{e^2}{\hbar c^3}\int \frac{d\omega\omega^2}{8\pi} 
\left[ \left(f_{\hbar\omega/2}^{(0)}\right)^2 +\frac{1}{2} \left(f_{\hbar\omega/2}^{(1)}\right)^2\right],\\
 \label{totrate-fin}
\end{eqnarray}
where
\begin{eqnarray}
f_{\hbar\omega/2}^{(0)} & = & \frac{1}{1+\exp\left(\frac{\hbar\omega}{2k_B T_e}\right)},\\  
 f_{\hbar\omega/2}^{(1)} & =  &  \frac{\mu {\cal E}_x}{v}\frac{\hbar\omega}{8k_B T_e\cosh^2\left(\frac{\hbar\omega}{4k_B T_e}\right)}.
\end{eqnarray}
Equation (\ref{totrate-fin}) is simply the spontaneous emission rate of quantum dipoles represented by electron-hole pairs in graphene.
The ratio between the SPP and photon emission rates is known as the Purcell factor shown in {\bf Fig. 1(f)} of the main text.

It is also instructive to compute the Purcell factor at non-zero quasi Fermi energy levels for electrons and holes
that may represent an additional THz excitation of graphene. There is no considerable effect though,
because increasing the number of electron-hole pairs proportionally enhances both photon and SPP emission rates.

Finally, we calculate the photon emission power that is needed for the total power balance including SPP emission power and electron (Joule) power.
The derivation follows the same route as for the photon emission rate (\ref{totrate}) with the integrand multiplied by $\hbar\omega$.
In addition, we consider emission not to vacuum but to a dielectric media with the refractive index $n_d$.
The results reads
\begin{eqnarray}
\nonumber P_{ph} & = & \frac{4}{3}\frac{n_d e^2}{\hbar c^3}\int \frac{d\omega\omega^3}{8\pi}
\left[ \left(f_{\hbar\omega/2}^{(0)}\right)^2 +\frac{1}{2} \left(f_{\hbar\omega/2}^{(1)}\right)^2\right].\\
 \label{totrate-power}
\end{eqnarray}
The integral can be written in terms of the $\zeta$-function, but we make use of equation (\ref{totrate-power}) as it is with $n_d\sim 2.2$ (average value for hBN)
to compute the power conversion efficiency shown in {\bf Fig. 4} of the main text.
\end{document}